\documentclass[sigconf]{acmart}

\usepackage[linesnumbered,ruled,vlined]{algorithm2e}
\usepackage{subcaption}
\usepackage{xcolor}
\usepackage{listings}

\AtBeginDocument{%
  \providecommand\BibTeX{{%
    \normalfont B\kern-0.5em{\scshape i\kern-0.25em b}\kern-0.8em\TeX}}}

\setcopyright{acmcopyright}
\copyrightyear{2018}
\acmYear{2018}
\acmDOI{XXXXXXX.XXXXXXX}

\copyrightyear{2025}
\acmYear{2025}
\setcopyright{rightsretained}
\acmConference[DIS '25]{Designing
Interactive Systems Conference}{July 05--09, 2025}{Madeira, Portugal}
\acmBooktitle{Designing
Interactive Systems Conference (DIS '25), July 05--09, 2025, Madeira, Portugal}
\acmDOI{}
\acmISBN{}

\newcommand{\cmmnt}[1]{}

\author{Xiyun Hu}
\orcid{0000-0001-9497-6925}
\authornote{Both authors contributed equally to this research.}
\affiliation{%
  \institution{School of Mechanical Engineering \\ Purdue University}
  \city{West Lafayette}
  \country{USA}}
\email{hu690@purdue.edu}

\author{Dizhi Ma}
\orcid{0009-0005-1711-4930}
\authornotemark[1]
\affiliation{%
  \institution{Elmore Family School of Electrical and Computer Engineering \\ Purdue University}
  \city{West Lafayette}
  \country{USA}}
\email{ma742@purdue.edu}

\author{Fengming He}
\orcid{0000-0002-8558-8848}
\affiliation{%
  \institution{Elmore Family School of Electrical and Computer Engineering \\ Purdue University}
  \streetaddress{610 Purdue Mall}
  \city{West Lafayette}
  \state{IN}
  \country{USA}
  \postcode{47907}
}
\email{he418@purdue.edu}

\author{Zhengzhe Zhu}
\orcid{0000-0001-9935-0518}
\affiliation{%
  \institution{Elmore Family School of Electrical and Computer Engineering \\ Purdue University}
  \city{West Lafayette}
  \country{USA}}
\email{zhu714@purdue.edu}

\author{Shao-Kang Hsia}
\affiliation{%
  \institution{School of Mechanical Engineering \\ Purdue University}
  \city{West Lafayette}
  \country{USA}}
\email{shsia@purdue.edu}

\author{Chenfei Zhu}
\orcid{0009-0003-3408-2876}
\affiliation{%
  \institution{School of Mechanical Engineering \\ Purdue University}
  \city{West Lafayette}
  \country{USA}}
\email{zhu1237@purdue.edu}

\author{Ziyi Liu}
\orcid{0000-0002-1270-2734}
\affiliation{%
  \institution{School of Mechanical Engineering \\ Purdue University}
  \city{West Lafayette}
  \country{USA}}
\email{liu1362@purdue.edu}

\author{Karthik Ramani}
\orcid{0000-0001-8639-5135}
\affiliation{%
  \institution{School of Mechanical Engineering \\ Purdue University}
  \city{West Lafayette}
  \country{USA}}
\email{ramani@purdue.edu}
\newif \ifdraft \drafttrue   

\newif \ifhighlight \highlightfalse    

\usepackage[normalem]{ulem}

\providecolor{added}{RGB}{65,105,225}   
\providecolor{deleted}{rgb}{1,0,0}     
\newcommand{\added}[1]{{\ifhighlight {{\color{added}{#1}}}\else{#1}\fi}}
\newcommand{\deleted}[1]{{\ifhighlight {{}}\fi}}

\begin{document}

\title[GesPrompt]{GesPrompt: Leveraging Co-Speech Gestures to Augment LLM-Based Interaction in Virtual Reality}


\renewcommand{\shortauthors}{Hu and Ma, et al.}
\begin{abstract}



























Large Language Model (LLM)-based copilots have shown great potential in Extended Reality (XR) applications. However, the user faces challenges when describing the 3D environments to the copilots due to the complexity of conveying spatial-temporal information through text or speech alone. To address this, we introduce \oursystem, \deleted{a copilot system} \added{a multimodal XR interface} that combines co-speech gestures with speech, allowing end-users to communicate more naturally and accurately with LLM-based copilots in XR environments. By incorporating gestures, \deleted{\oursystem~improves the copilot's ability to understand spatial references} \added{\oursystem~extracts spatial-temporal reference from co-speech gestures}, reducing the need for precise textual prompts and minimizing cognitive load for end-users.  Our contributions include (1) a workflow to integrate gesture and speech input in the XR environment, (2) a prototype VR system that implements the workflow, and (3) a user study demonstrating its effectiveness in improving user communication in VR environments.

\end{abstract}




\received{20 January 2025}
\received[Conditionally Accepted]{8 April 2025}
\received[Revised]{21 April 2025}


\newcommand{\oursystem}{GesPrompt}
\newcommand{\bfit}[2][black]{\textbf{\textit{\textcolor{#1}{#2}}}}

\maketitle

\section{Introduction}
\begin{figure*}
\centering
  \includegraphics[width=\textwidth]{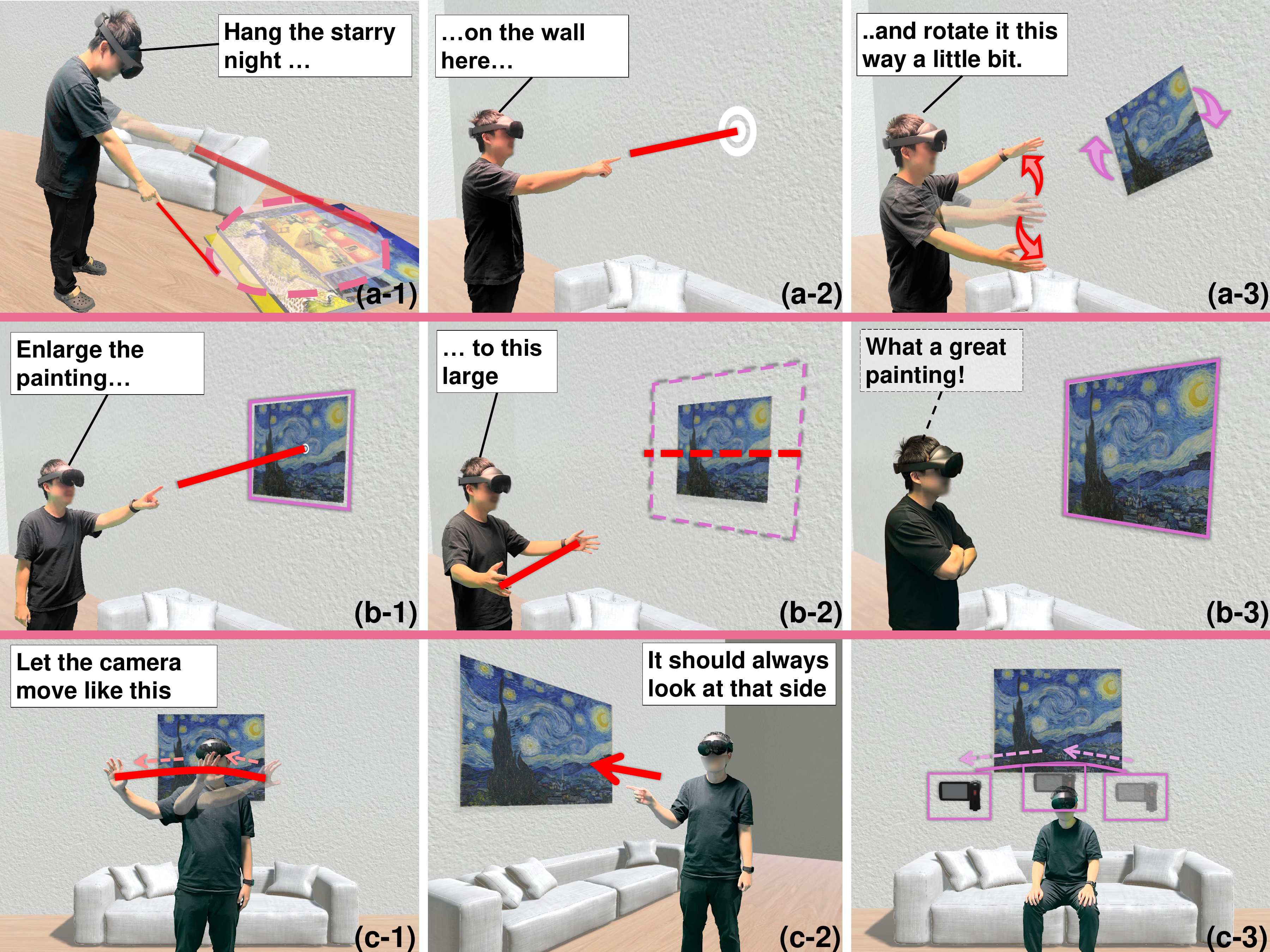}
  \caption{An example usage of \textit{\oursystem}: (a) An end-user wants to hang a virtual painting on the wall using \textit{\oursystem}: (a-1) The user asks the system to select the ``Starry Night'' painting from a pile of paintings on the floor by gesturing the rough location of the painting; (a-2) in the same utterance, the user also specifies the location of the painting using gesture; (a-3) realizing that the painting is not level, the user instructs the system to adjust its alignment while gesturing the appropriate amount of rotation; (b) The user stands back and sees that the painting is too small, then asks the system to enlarge the painting:(b-1) (b-2) the user sets the size of the painting by gesturing; (b-3) the user is satisfied with the painting, now properly placed and resized to fit the room's aesthetics. (c) Then the user is excited and wants to take a virtual video with the painting: (c-1) The user asks the system to set up the path of the camera by gesturing; (c-2) they also want the camera to look at a certain direction, which is pointed out by the user; (c-3) the camera moves along the path and records the video as the user desires.
}
  \label{fig:teaser}
  \Description{Nine panels depict a person wearing a VR headset using speech and hand gestures to manipulate a virtual “Starry Night” painting in a living‑room scene. The first row shows hanging the painting: first, the user points at the painting on the floor; next, they point to a spot on the wall; finally, they twist both hands as the painting floats and rotates into position. The second row shows resizing the painting: first, the user points at the mounted painting; then they spread their hands under it to indicate a larger size; then they step back with arms crossed, admiring the now‑enlarged painting. The third row illustrates camera control: first, the user sweeps both hands side to side to define a camera path; next, they point at the painting to set the camera’s viewing angle; finally, they sit on a sofa while a camera icon moves along the sketched path toward the painting.
}
\end{figure*}

With the development of Large Language Models (LLMs), LLM-based copilots on desktops and smartphones that assist end-users with auto-generated suggestions received a wide range of recognition in various domains, such as coding \cite{github2023copilot}, writing~\cite{openai2023chatgpt}, and day planning \cite{microsoftplanner}. 
Typically, these applications leverage the textual context understanding ability of the LLM to understand the end-user's intention and offer assistance to fulfill the user's need.
Recently, researchers have explored integrating LLM-based agents into extended reality (XR) environments for various applications, such as social games~\cite{wan2024building}, education \cite{liu2024classmeta,chheang2024towards}, and spatial design \cite{zhang2024vrcopilot}. 
Although the above-mentioned LLM-based copilots show great potential in assisting users in XR, interacting with 3D environments using \added{text-only or speech-only} prompts remains challenging for end-users \added{to express complex spatial–temporal intents}.

\deleted{First, current LLM-based copilots have demanding requirements for end-users. 
Existing copilots usually require end-users to provide simple yet carefully engineered prompts or step-by-step instructions to ensure that the LLMs can accurately understand users' intentions. Consequently, novice users often have to make multiple attempts to achieve their desired results. }
\added{First, end-users must craft carefully engineered prompts or precise descriptions to ensure accurate LLM interpretation, often resulting in trial-and-error interactions.}
Second, tasks in 3D environments involve intricate spatial-temporal information, and such information is difficult to describe precisely and efficiently in natural language. Most copilots that focus on text/speech input struggle to enable end-users to accurately convey complex spatial-temporal details, as these text/speech-only inputs are prone to pronoun ambiguity \cite{aikawa2023introducing, zhang2024vrcopilot, lee2024gazepointar}. 
For example, an end-user wants to refer to an unfamiliar object in a 3D environment to a text-only copilot, then the user has no choice but to describe the physical characteristics of the object in detail.
\added{This multi-step prompting increases cognitive load and disrupts workflow~\cite{ping2010gesturing, broaders2010truth}. Consequently, enabling effortless interaction for complex 3D tasks in XR remains an open challenge.}
\deleted{This process becomes cumbersome for the user when they need to handle multiple prompts, leading to an increased mental burden and cognitive load} 
\deleted{ \cite{broaders2010truth}\cite{ping2010gesturing}.}
\deleted{Hence, how to empower end-users to effortlessly use copilots for complex 3D tasks in XR environments remains an open problem in LLM deployment research.}

Some prior works address the problem of end-users' difficulty in describing spatial-temporal information by incorporating scene understanding.
In these efforts, researchers guide end-users to describe the spatial-temporal information in an XR environment from high-level to low-level details \cite{de2024llmr, wu2022ai}. 
While these methods help and enable users to craft accurate prompts that are understandable by LLMs, the sheer number of users' prompts remains demanding. 
This results in a lengthy process that is cumbersome for end-users to complete. 
\deleted{On the other hand, combining LLMs with other human input, such as hand gestures and gazes, seems a plausible solution. In particular, gestures show great potential for augmenting LLM understanding in the 3D scenes. }
\added{On the other hand, multimodal input—combining LLMs with human modalities such as gesture and gaze—offers a promising path forward.}
Gestures, as a natural interaction modality, have long been adopted for interacting with 3D environments in HCI research~\cite{rautaray2015vision, guo2021human, yang2019gesture, shi2023hci}. 
Specifically, end-users can use various gestures such as grabbing~\cite{yousefi20163d, yan2018virtualgrasp}, touching~\cite{arora2019magicalhands,price2021conceptualising}, and air-taping~\cite{huang2015ubii, mendes2014mid, ong2011augmented} to manipulate objects in XR scenarios. 
Furthermore, compared with speech-only communication, gestures can convey essential non-verbal information that adds emphasis and clarity \cite{burgoon1990nonverbal, ao2022rhythmic, studdert1994hand, alibali2005gesture,bolt1980put}. 
For example, end-users may want to use gestures to illustrate the desired size of a virtual object when describing it to the copilot.
\deleted{Although gestures have shown great potential, current gesture applications in LLM copilots are still limited, with most work focusing on simple pointing gestures \cite{zhang2024vrcopilot}\cite{ lee2024gazepointar}.}
\added{Recent XR systems (e.g., GazePointAR~\cite{lee2024gazepointar}, VRCopilot~\cite{zhang2024vrcopilot}) leverage pointing alongside speech but remain limited to simple deictic input. Prior HCI work such as Grasp \& Speech~\cite{piumsomboon2014grasp}, Mixed Speech–Gesture Interfaces~\cite{friedrich2021combining}, and Multimodal Command Tools~\cite{chen2017multimodal} typically follow a two‑stage pattern: users first perform a manipulative gesture (e.g., pinch, drag) to select or highlight a virtual object, then issue a spoken command (e.g., “rotate,” “delete”) to trigger an action. While this sequential coupling supports straightforward object manipulation, it treats speech and gesture as largely independent phases and fails to exploit the rich semantic alignment that naturally occurs in co‑speech gestures.
Thus, we harness \emph{co‑speech gestures}—gestures naturally produced alongside spoken utterances—to fill gaps in verbal references and reduce prompting overhead.}

\deleted{We propose \oursystem, an LLM-powered XR copilot system that allows users to express their intention through the combination of speech and gestures, without the need to create engineered commands or mentally translate spatial-temporal information into numerical values.}
\added{We propose \oursystem, a multimodal XR interface powered by an LLM that allows users to express their intentions through integrated speech and co‑speech gestures, without requiring engineered commands or the mental translation of spatial‑temporal information into numerical values.}
The \oursystem~ workflow consists of two interconnected main components: 
(1) an LLM system that possesses both object information and the XR system function information. 
The LLM system takes the user's speech and converts it into a sequence of function calls that can be executed by the back-end XR system. 
Additionally, the LLM system identifies ambiguous phrases in the speech that cannot be understood through speech alone and sends them to the gesture processor. 
(2) a gesture processor that processes gesture inputs. 
The processor converts raw skeleton gesture data into meaningful parameter values \added{(position, object, direction, rotation, size, and path)} using cues, which are discovered by the LLM system, from speech. 
\deleted{Unlike conventional manipulative gestures in XR settings, co-speech gestures occur alongside speech, and their meaning can only be determined in the context of speech. To determine how to process gesture data, we identified six major types of spatial-temporal information that gestures can represent: position, object, direction, rotation, size, and path. }
\added{By jointly reasoning over speech and gesture, ~\oursystem~ enables intuitive, low-effort
interaction with 3D scenes.}

In summary, we highlight our contributions as: 
\begin{itemize} 
    \deleted{\item A system that is capable of processing natural voice and gesture input in XR environments.  }
    \item  \added{A novel co-speech gesture integration method for LLM-powered XR copilots, extending beyond simple deictic pointing.}
    \item A prototype VR copilot that implements the workflow for object manipulation using voice and gesture.
    \item A user study to evaluate the feasibility of the workflow and investigate user behavior when interacting with the system.
\end{itemize}

\section{Related Works}
\subsection{LLM-based Copilots for 3D Scenes}
As LLMs have emerged recently and gained widespread attention \cite{ touvron2023llama, achiam2023gpt, brown2020language, team2023gemini}, LLM-based copilots are becoming a convenient way for end-users without coding expertise to interact and communicate in the HCI area. With the assistance of copilots, end-users can easily create customized digital contents, including images \cite{brown2020language, strobelt2022interactive, wu2022ai}, UI designs \cite{kim2022stylette,  peng2024designprompt},  and 3D models/scenes \cite{jain2022zero, poole2022dreamfusion, chen2024vp3d, liu2024make, hong20233d, wang2023chat}, using simple prompts.  

Notably, significant advances in generative models for 3D scenes have captured HCI researchers' attention and motivated them to explore the integration of LLM-based copilots into interactive 3D environments. LumiMood \cite{oh2024lumimood} proposed a system that leverages LLMs to automatically adjust lighting and post-processing in 3D scenes, allowing the mood of scenes to dynamically adapt accordingly. Bozkir et al. \cite{bozkir2024embedding} and Zhu et al. \cite{zhu2023free} forecast that integrating LLM into XR as narrative agents or visual avatars with different strategies will enhance users' engagement in XR. Aikawa et al. \cite{aikawa2023introducing} propose an LLM-based system that assists end-users in brainstorming within the AR space. 
Meanwhile, MagicItem \cite{kurai2024magicitem} and DreamCodeVR \cite{giunchi2024dreamcodevr} enable end-users without coding expertise to prototype VR content's behavior. By utilizing LLM-based copilots to generate code within VR space, these two works empower users to program basic behavior of 3D contents, such as ``move'', ``jump'', and ``rotate''. Yet, the programmed behaviors are limited in scope as end-users are incapable of modifying more detailed aspects of the virtual content, such as altering the distance of movement. Besides LLM-based copilots, most state-of-the-art copilots such as GPT-4 \cite{achiam2023gpt} apply multimodal models, which accept different sources of inputs including texts and images.  VR-GPT \cite{konenkov2024vr} provides end-users a natural way to complete object collection tasks by interacting with a Vision Language Model (VLM)-based copilot, which is capable of processing both images and language prompts within VR environments. 

While the aforementioned works have received recognition for applying robust LLM-based copilots in XR, most of the works focus on entry-level design and prototyping for end-users. Consequently, end-users remain incapable of performing complicated 3D tasks involving detailed spatial-temporal information or dynamic interactions, such as precise movements of virtual content or responsive adjustments according to users' status. 
To let end-users accurately generate prompts in interactive XR environments, recent work such as LLMR \cite{de2024llmr} that incorporates scene understanding breaks down complex 3D tasks into several subtasks and employs an ensemble of multiple language models, each addressing a specific task. While this approach enhances the robustness of language models in XR environments, the sheer number of subtasks and associated prompt numbers makes the overall progress overwhelming and time-consuming for end-users. Motivated by these challenges, we strive to develop a system that streamlines the process of prompting LLM-based copilots for complex 3D tasks and interactions, so that end-users can effortlessly utilize these copilots in XR environments.




\subsection{Gestures in XR Multimodal Interaction}

Gestures, being a natural and intuitive human input, have long been recognized in the HCI area for facilitating interactions with 3D environments in daily activities \cite{wang2021gesturar, arora2019magicalhands, buchmann2004fingartips, zhou2020gripmarks, lu2012gesture, yousefi20163d}. Through gestures, end-users can effortlessly manipulate 3D objects via body or hand movements. For example, end-users can easily use simple gestures, such as pointing \cite{lee20083d}, pinching \cite{huang2015ubii, mendes2014mid}, and grabbing \cite{zhou2020gripmarks, yousefi20163d} to select or move 3D content in XR environments. Beyond basic interactions, gestures also enable end-users to modify the complex behavior of virtual content. MagicHands \cite{arora2019magicalhands} empowers end users to perform customized gestures to author particle animations. GesturAR \cite{wang2021gesturar} enables end-users without coding expertise to prototype AR applications and modify AR content behaviors through customized freehand gestures. ProGesAR \cite{ye2022progesar} integrates gestures with IoT devices, allowing end-users to trigger IoT functions through customized gestures. Besides providing intuitive interactions, gestures also allow end-users to effectively convey spatial-temporal information in XR environments \cite{chen2012kinetre, nishino19983d}.  SketchingWithHands \cite{kim2016sketchingwithhands} allows users to use gestures to describe hand-held objects' shape or contour. KinÊtre \cite{chen2012kinetre} introduces a system that allows virtual content to change poses by mimicking end-users' postures in the 3D environment. 

\added{
Combining gestures with speech offers a powerful multimodal approach. Early interfaces (e.g., Grasp \& Speech \cite{piumsomboon2014grasp}, Freehand–Voice Manipulation \cite{lee2013usability}, Mixed Speech–Gesture Interfaces \cite{friedrich2021combining}, and Multimodal Interaction with Virtual Agents \cite{chen2017multimodal}) follow a two‑phase pattern: users first perform a gesture to designate or highlight an object, then issue a spoken command (e.g., “rotate,” “delete”) to specify the action. While effective for discrete tasks, this sequential coupling treats speech and gesture as independent stages and relies on engineered command vocabularies.

}
Recently, researchers attempted to leverage the intuitiveness and expressiveness of gestures to assist end-users in overcoming the challenges of using LLM-based copilots for 3D tasks in XR environments. When end-users communicate with copilots in 3D environments, one major issue is the pronoun ambiguity for describing spatial-temporal information
\cite{aikawa2023introducing, zhang2024vrcopilot, lee2024gazepointar, aghel2024people}. Aghel et al. \cite{aghel2024people} present a Wizard-of-Oz elicitation study and highlight the importance of using gestures for reference when end-users generate prompts in 3D environments. Further, Wu et al. \cite{wu2024body} investigate when and how full-body gestures can be used together with LLM-based copilots. Targeting the problem of pronoun ambiguity, GazePointAR \cite{lee2024gazepointar} enables end-users to leverage eye gaze as well as pointing gestures to describe specific locations in 3D scenes when talking with copilots.  Similarly, VRCopilot \cite{zhang2024vrcopilot} combines speech and pointing gestures to allow end-users to specify their creation needs in immersive environments. 
\deleted{However, most of these systems primarily use gestures to indicate the location of 3D content, overlooking other important spatio-temporal information such as rotation, size/shape changes, and object movement. 

While we acknowledge the expressiveness and intuitiveness of gestures in XR interactions, current applications of gestures in LLM-based copilots are limited to simple pointing gestures and therefore remain limited in conveying the full scope of spatio-temporal information. To fill this gap, our system is designed to support end-users in effortlessly communicating their intentions and accurately conveying spatio-temporal details with an LLM-based copilot, fully leveraging the characteristics of gestures.}
\added{However, these approaches support only simple location cues and do not leverage the full expressiveness of co‑speech gestures, which naturally accompany spoken utterances and can encode rotation, size, and motion trajectories.

To address this gap, we leverage \emph{co‑speech gestures}---gestures produced concurrently with speech---to enrich LLM understanding of spatial-temporal intents. By jointly interpreting speech and gesture in our multi‑modal XR interface, \oursystem~ reduces the need for carefully engineered prompts or mental translation of spatial data into numerical values, enabling fluid and intuitive interactions with 3D scenes.}

\subsection{Synergy Between Speech and Gesture: Co-speech Gestures}
\label{sec:r3}

In human communication, co-speech gestures can provide additional context, complementing speech to communicate intention more clearly \cite{clough2020role, kelly2023exploring}, especially regarding spatial and physical aspects \cite{wu2014co,ping2010gesturing}. 
Following the definition from a pioneering work - Hand and Mind \cite{studdert1994hand}, co-speech gestures can be categorized into four types: deictic gestures, iconic gestures, metaphoric gestures, and beats. This classification is widely adopted in the design and analysis of co-speech gestures \cite{diessel2020demonstratives, bolt1980put, willems2007language, kita2003does, masson2017we}. 

\textbf{Deictic gestures}, often known as the pointing gesture \cite{diessel2020demonstratives}, are used to refer to entities in the environment. The pioneer work Put-that-there \cite{bolt1980put} in the HCI area can serve as an example: the user says ``put that there'' while pointing to an object on screen and pointing to another location, adding missing information from the speech.

\textbf{Iconic gestures} are used to depict the spatial or action aspect of the elements in the speech \cite{willems2007language,kita2003does,masson2017we}. For example, when someone says ``you cannot believe the huge bird I saw today'', they may spread both arms wide apart to visually indicate a large size.

\textbf{Metaphoric gestures} are similar to iconic gestures. They both describe the physical aspect of a concept, but the difference is that iconic gestures illustrate concrete contents while metaphoric gestures is used to present abstract concepts \cite{straube2011differentiation}. For example, the spreading arm gesture can also demonstrate the level of welcoming, when someone says ``I welcome you with my whole heart.''

\textbf{Beats} are often biphasic gestures to emphasize some words or phrases in a sentence \cite{masson2017we}. For example, when someone says ``you have to be there \underline{now}'' with a downbeat gesture, they mean to emphasize the timing.  

Among the above gestures, we mainly focus on deictic gestures and iconic gestures due to their expressiveness in representing the spatial-temporal information.

In order to extract the spatial-temporal information from the deictic gestures and iconic gestures, the detection of such gestures is the first step.
Deictic gestures, which convey only the meaning of referring to something, are usually static and can be detected with current gesture recognition models ~\cite{kopuklu2019real, madhiarasan2022comprehensive}. In commercial systems such as HoloLens 2 \cite{hololens} and Meta Quest Pro \cite{Oculus}, a ray is constantly cast from the center of the user's hand, and the user can perform a pinch gesture to confirm the selection or reference of objects or entities. 
The iconic gesture, on the other hand, can represent the action aspect of an object, making the gesture sometimes dynamic. 
Moreover, the wide variety of shapes and actions that the iconic can represent make it even harder for the traditional gesture detection model to recognize \cite{sowa2001interpretation}.
Recent work by Ghaleb et al. ~\cite{ghaleb2024leveraging} explores a framework that leverages speech information in detecting gestures.  
Zhang et al. \cite{zhang2024semantic} find it promising to use LLM in aid of synthesizing co-speech gestures. 
Building on these insights, and the fact that co-speech gesture will occur when the user says the word that is relevant to it ~\cite{holle2008neural,wagner2014gesture}, our system utilizes speech to detect the appearance of the co-speech gesture, and analyzes the co-speech gesture based on the context given in speech. 

Although deictic and iconic gestures are effective in representing certain actions or spatial relationships, they naturally have limitations in conveying more complex or abstract meanings.
Speech, which evolved in part from gestures, complements these limitations by covering the meanings that gesture alone cannot easily express. 
This synergy between speech and gesture motivates our design of a system that integrates both inputs, allowing speech to guide the interpretation of gestures, and gesture to complete the missing information from speech.


\section{\oursystem~}
\label{sec:s3}

In this work, we aim to let end-users communicate with LLM-based copilots while conveying spatial-temporal information in XR environments. To achieve this, we design a workflow that processes natural speech and gesture input from users simultaneously. 
When a user starts a speech and provides gesture command to \oursystem, the system follows the following steps: (1) extracts users' inputs, including speech and gestures, (2) interprets user input by referencing appropriate virtual content and behaviors, and (3) determines the backend \bfit{functions} ($F$) responsible for controlling targeted behaviors.

Here, we formally define the process as follows:
Let $F$ represent a set of available system \textit{functions} $F_1,F_2,...,F_n$, and let \textbf{$S$} denote the current state of the XR environment. Calling these functions will alter the environment's state $S$. Each function $F_i$ takes a set of \bfit{function parameters} $P_i=\{p_1,p_2,...,p_k\}$, where $p_i$ represents properties of the XR environment such as objects and spatial positions.

Given users inputs of (1) \textbf{\textit{text input}} $T$: verbal descriptions of the desired state of the XR environment, and (2) \textbf{\textit{gestural input}} $G$: gestural cues that provide additional context to the \textit{text input}, such as hand movements for outlining the shape of an object in the environment,
our goal is to:
\begin{itemize}
    \item Infer the correct sequence of function calls
    
    $C=\{(F_{i1},P_{i1}),(F_{i2},P_{i2}),...,(F_{im},P_{im})\}$ from the \textit{text input}.
    \item Extract the appropriate values for the \textit{function parameters} $P_{im}$ from both text and gesture input.
    \item Apply the function calls $ C$ to transition the XR environment state: $S_{current} \xrightarrow[]{C} S_{intended}$.
\end{itemize}

In the following sections, we describe how \oursystem~ achieves the goal and its detailed design.

% ...

%
\subsection{System Walk-through}

\begin{figure*}[h]
    \centering
    \includegraphics[width=\linewidth]{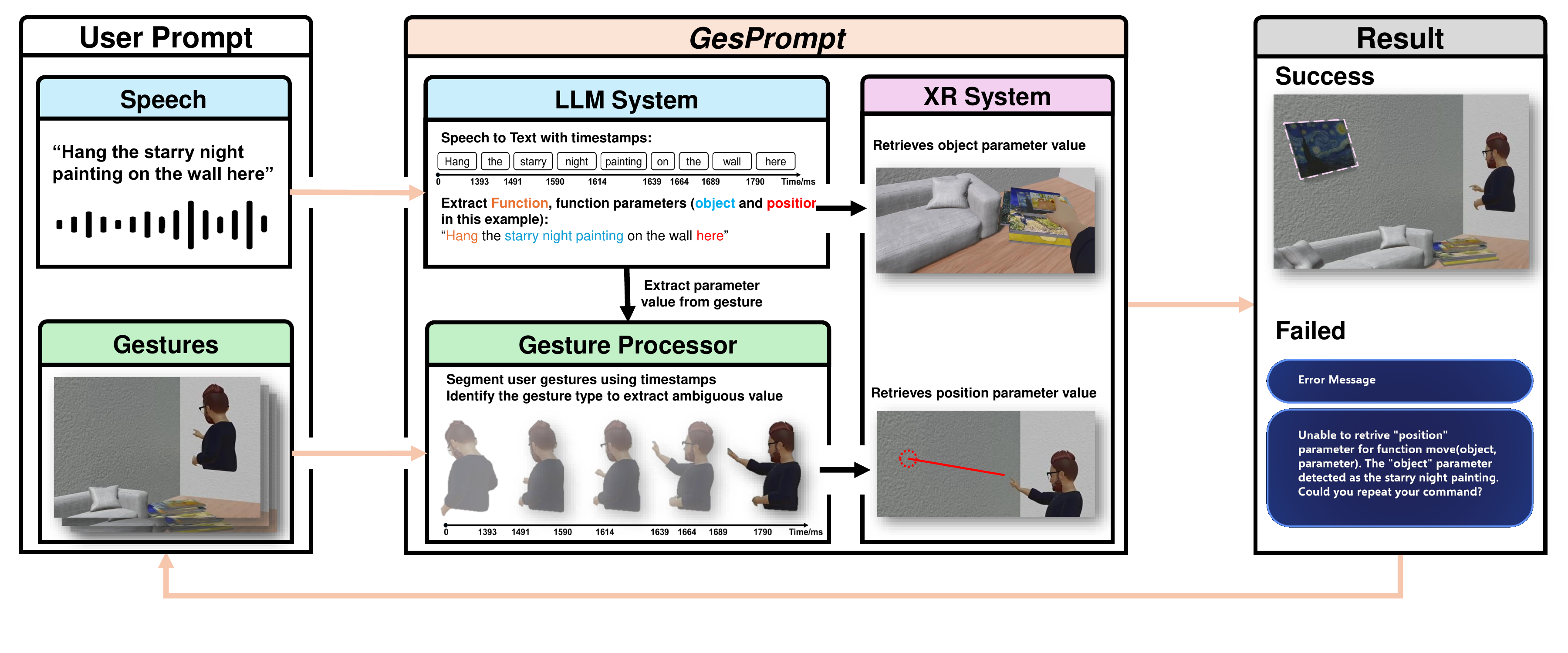}
    \caption
    {
    System Workflow: The system modifies the XR environment using speech and gesture prompts, processed by two core components: the LLM system and the Gesture Processor.    
    The LLM system converts speech to text with timestamps, identifies operation functions and parameters, and extracts unambiguous values from the XR scene. 
    The Gesture Processor resolves ambiguous parameters by analyzing gestures segmented using speech timestamps and extracting values from the XR scene. 
    Once all parameters are identified, the system updates the XR environment.
    If the system fails to update the environment, an error message will be sent to the user.
    }
    \Description{Diagram showing the end‑to‑end data flow in GesPrompt: Leftmost panel titled “User Prompt” contains two sections. The top shows a speech waveform icon and the text “Hang the Starry Night painting on the wall here.” The bottom shows a small VR scene screenshot of a user in a headset pointing at a painting on the floor and at a wall. The center panel titled “GesPrompt” is divided into three subpanels. The top‑left subpanel labeled “LLM System” shows a transcription of the speech broken into words with time‑stamp markers underneath, and the extracted function call indicating the parameters “object” and “position.” The bottom‑left subpanel labeled “Gesture Processor” displays a timeline of silhouette thumbnails of the user’s hand pointing gesture at the corresponding time stamps, with the caption about segmenting and identifying ambiguous parameters. On the right side of the center panel is a subpanel labeled “XR System” showing two screenshots: one of the user’s hand selecting the painting object, and one of the user pointing at the wall to retrieve the position parameter. The rightmost panel, titled “Result” shows two outcomes. Under “Success” is a VR scene where the painting now hangs on the wall and a virtual avatar looks on. Under “Failed” is a stylized error dialog reading “Unable to retrieve ‘position’ parameter for function move(object, position). The ‘object’ parameter was detected as the Starry Night painting. Could you repeat your command?” An arrow loops back from the Result panel to the User Prompt panel to indicate retry on failure.}
    \label{fig:Framework}
\end{figure*}

We introduce \oursystem, a workflow designed for users to interact with LLM-based copilots through speech and gesture. As illustrated in \autoref{fig:Framework}, this workflow is composed of three main components: the \textbf{\textit{LLM System}}, the \textbf{\textit{Gesture Processor}}, and the \textbf{\textit{XR system}}. The workflow of GesPrompt is demonstrated with an example scenario in \autoref{fig:Framework}.

An end-user wants to hang a painting on the wall but struggles to verbally describe the exact position of the painting. With our system, which continuously listens to the user's command, the user simply says ``Hang the Starry Night painting'' while pointing to the paintings, then says ``...on the wall here'' while pointing to the wall. \oursystem~ handles the speech and gesture input with the following steps: first, the \textit{LLM System} identifies the appropriate backend function (\(F\)),  for the task from the speech, i.e., selecting \(F_{move}(object, position)\) for this scenario to place the painting on the wall.  Then, the LLM System maps the \textbf{\textit{parameter tokens}} from the speech to \textit{function parameters}. In this scenario, ``Starry Night painting'' \added{(object name)} is mapped to the \textit{object parameter}  \added{(a virtual object in the scene)}, and ``here'' is mapped to the \textit{position parameter}. With the scene information \added{(e.g., scene objects' names, positions, rotations, and scales)} fed into the LLM System, the target painting can be identified through the object parameter. However, the position parameter ``here'' is identified as an \textbf{\textit{ambiguous parameter}} (\(P_{i}^{amb}\)), as its value cannot be determined. 

To resolve this ambiguity, \oursystem~ forwards the ambiguous parameter and the corresponding \textit{timestamps} of its parameter token to the \textit{Gesture Processor} for further analysis. In this case, a position parameter and the timestamps associated with the user’s utterance of “here” are sent to the Gesture Processor. The gesture that occurred during the interval is then segmented. Since the parameter type is ``position'', the segmented gesture is identified as a pointing gesture, and the position where the user points is further used as the exact position parameter.

Finally, \oursystem~ checks for any invalid or incomplete functions and parameter values in the \textit{XR System}. If all functions and values are valid, the XR System then executes the \textit{function calls} (\(C\)), transitioning the XR environment state to the user’s intended state. In this case, the Starry Night painting is moved to the position where the user points. When the system detects missing or invalid parameters for the target functions, a textbox will pop up prompting the user to provide further clarification.
In the following sections, we describe the design of each component of \oursystem.

\subsection{Spatial-Temporal Parameter and Co-Speech Gestures}


Although it’s possible to articulate spatial-temporal parameters through speech alone, integrating co-speech gestures often provides a more efficient and intuitive means of conveying these parameters. In this section, we concentrate on scenarios where users employ co-speech gestures to depict spatial-temporal parameters. As discussed in Section \ref{sec:r3}, co-speech gestures are context-dependent. A single gesture may convey different meanings based on the accompanying speech. We summarize the characteristics of co-speech gestures and their representations of spatial-temporal parameters in the context of an extended reality environment.

\subsubsection{Deictic Gestures}

\begin{figure}[h]
    \centering
    \includegraphics[width=\linewidth]{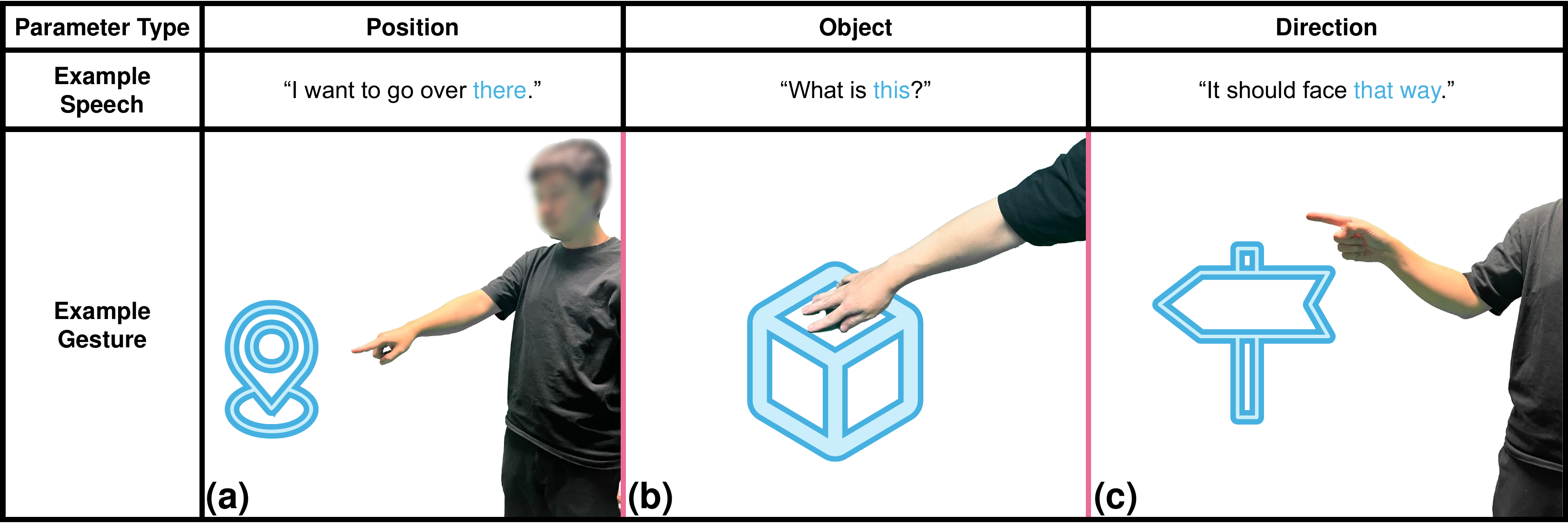}
    \caption{Example showcase of deictic gestures and the spatial parameters. 
    For the Position parameter, the gesture could be pointing toward a specific location. 
    For the Object parameter, the gesture could be pointing directly at an object. 
    For the Direction parameter, the gesture could be a directional pointing motion to indicate the desired orientation
    }
    \Description{A three‑column table illustrating three spatial parameter types—Position, Object, and Direction—with example speech and corresponding gestures. Under “Position,” the example speech reads “I want to go over there.” Below it is a location‑pin icon alongside a photograph of a person pointing straight ahead. Under “Object,” the speech says “What is this?” Below it is a 3D cube icon with a hand hovering over and touching its top face. Under “Direction,” the speech reads “It should face that way.” Below it is a signpost icon alongside a photograph of a person pointing to the right. Each column is labeled (a), (b), and (c) respectively.}
    \label{fig:deictic_table}
\end{figure}


This type of gesture indicates a location, entity, or direction within the environment. In a natural manner, index-finger-pointing is commonly adopted for entity selection, while whole-hand-pointing can be used in situations that require less precision~\cite{cochet2014deictic}. At the same time, the pointing gesture also varies in different cultures. To accommodate variations in pointing deictic gestures, we visualize the user's pointing direction by constantly casting a ray from their hand. This visualization, which serves as guidance for users to indicate desired entities, is a common practice in virtual and augmented reality applications~\cite{meta-sdk,microsoft_mrtk}.

As shown in \autoref{fig:deictic_table}, the same deictic gesture can represent three types of spatial parameters under different speech contexts: (a) position, (b) object, and (c) direction.


\textbf{Position} refers to a 3D world coordinate. The user may point to indicate a specific location (\autoref{fig:deictic_table}(a)). The intersection point between the pointed location and the ray cast from the user's hand provides the value for the position parameter. If the intended location is obscured by other objects, our system uses the user’s speech context to disregard any mentioned obstacles. For instance, when organizing a virtual room, the user wants to move a chair behind a table in front of the user.  If the user points behind the table and says ``The chair should go there, behind the table.'', the system interprets the table’s presence and places the chair in the correct position.

\textbf{Object} refers to an entity in the environment, such as virtual contents, humans, or part of an object. The user may point to or directly touch an entity to refer to it (\autoref{fig:deictic_table}(b)). We apply a similar method to process the object parameter as we do for the position parameter, with some variations. Instead of returning coordinates, the object that intersects with the ray is the output. Selecting an object requires less precision compared to setting the position parameter because an object's region typically encompasses a larger area than a precise location. Therefore, the touch gesture depicted in \autoref{fig:deictic_table}(b) can be interpreted as an up-close pointing gesture intended to select the object being touched by the user.



\textbf{Direction} refers to a spatial orientation or a directional vector in the environment. By pointing within the environment (\autoref{fig:deictic_table}(c)), the user specifies a direction. The directional vector of the ray cast from the user's hand provides the value for this parameter.

\subsubsection{Iconic Gestures}

\begin{figure}[h]
    \centering
    \includegraphics[width=\linewidth]{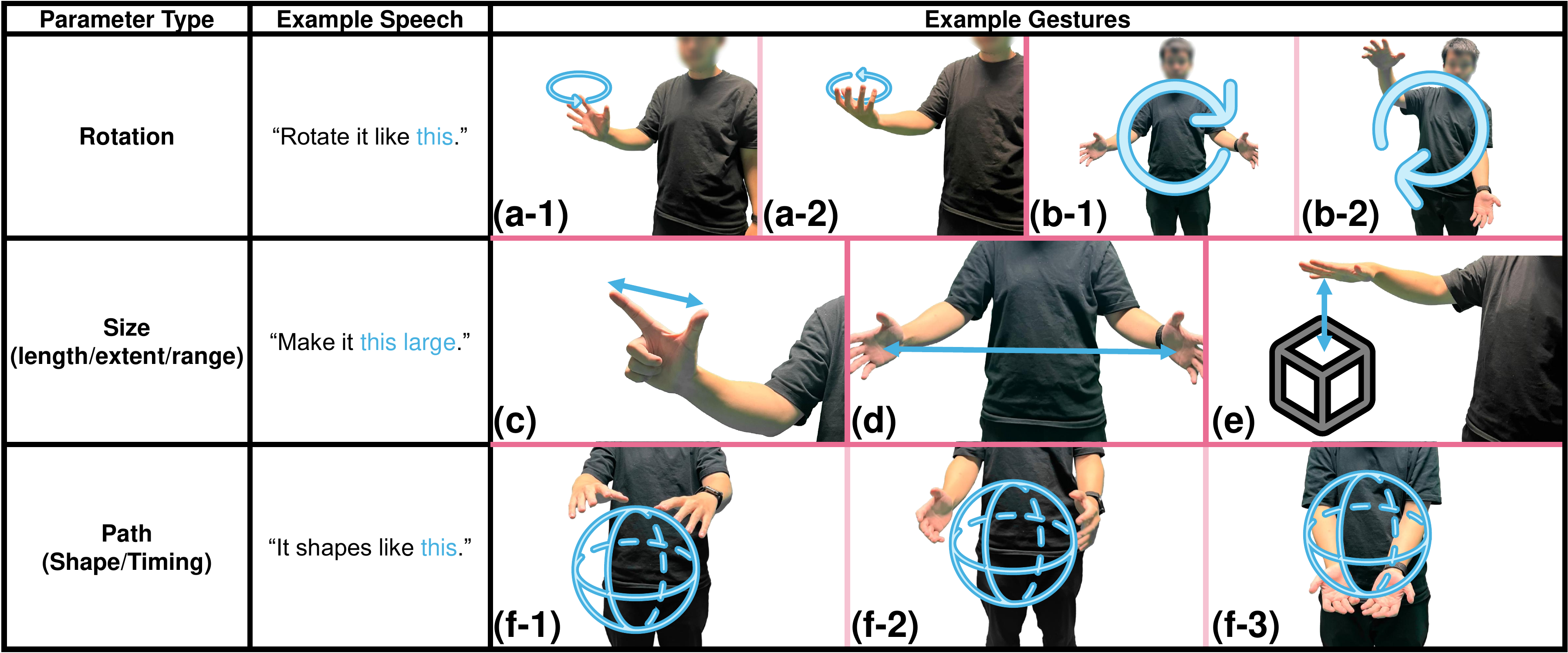}
    \caption{Examples of iconic gestures and the corresponding spatial parameters. 
    For the Rotation parameter, gestures could include the rotation of one (a) or both (b) hands to represent rotational movements. 
    For the Size parameter, gestures could indicate the relative size of an object using the distance between fingers (c), hands (d), or between a hand and a reference entity (e). 
    For the Path parameter, gestures could involve circular hand motions (f) outlining the shape of a sphere or demonstrating the trajectory of an object.
    }
    \Description{A table with three rows for Rotation, Size, and Path parameters, each showing example speech and co‑speech gestures: the Rotation row has the speech “Rotate it like this.” alongside two gesture thumbnails—first a single‑hand circular motion labeled (a‑1) and then a two‑hand circular motion labeled (a‑2), and two corresponding icon overlays labeled (b‑1) and (b‑2); the Size row shows “Make it this large.” with three gesture thumbnails—thumb‑index pinch labeled (c), two‑hand stretch labeled (d), and a hand above a cube icon indicating height labeled (e); the Path row shows “It shapes like this.” with three gesture thumbnails—hands drawing a spherical outline labeled (f‑1), a more complete sphere labeled (f‑2), and a fully sketched globe‑like path labeled (f‑3).}
    \label{fig:iconic_table}
\end{figure}

Instead of referencing external entities, the user can also refer to the hand gestures themselves through iconic gestures to express spatial parameters. We summarized three categories of spatial parameters that can be represented by iconic gestures, as illustrated in the examples in \autoref{fig:iconic_table}: (a) Rotation, (b) Size, (c) Path.

\textbf{Rotation} also refers to spatial orientation, but in this case, the gesture mimics the full rotation process.
For this parameter, there are two cases: the user imagines holding and rotating the object with one hand (\autoref{fig:iconic_table}(a)) or two (\autoref{fig:iconic_table}(b)). 
By monitoring changes in the position of the user's hands during the gesture, we can determine whether one or both hands are involved. 
If only one hand is used, the palm's rotation change during the gesture will be the output. 
If both hands are involved, we calculate the rotation change based on the line formed between the hands.

\textbf{Size} refers to the dimension or scale of an object. Deriving from it, length, extent, and degree parameters share the same characteristics as the size parameter, as they all relate to measurable aspects of entities. 
The size parameter can be represented using three types of gesture: using the distance between fingertips (i.e. thumb and index) (\autoref{fig:iconic_table}(c)); using the distances between hands (\autoref{fig:iconic_table}(d)); using the distance between the hand and another entity (e.g. hand hovering over the tabletop surface to indicate height) 
(\autoref{fig:iconic_table}(e)).
We calculate the distance accordingly using the gesture data, and the resulting distance becomes the value for this parameter. 


\textbf{Path} refers to a series of positions, rotations, and timestamps. By demonstrating a path, the user can sketch a shape in mid-air using both hands, as shown in \autoref{fig:iconic_table}(f). Alternatively, the user can generate a trajectory for a camera movement, as depicted in \autoref{fig:teaser}(c). The sequence of hand joint positions and rotations will be the value for the path parameter.

\subsection{LLM System}

\begin{figure}[h]
    \centering
    \includegraphics[width=\linewidth]{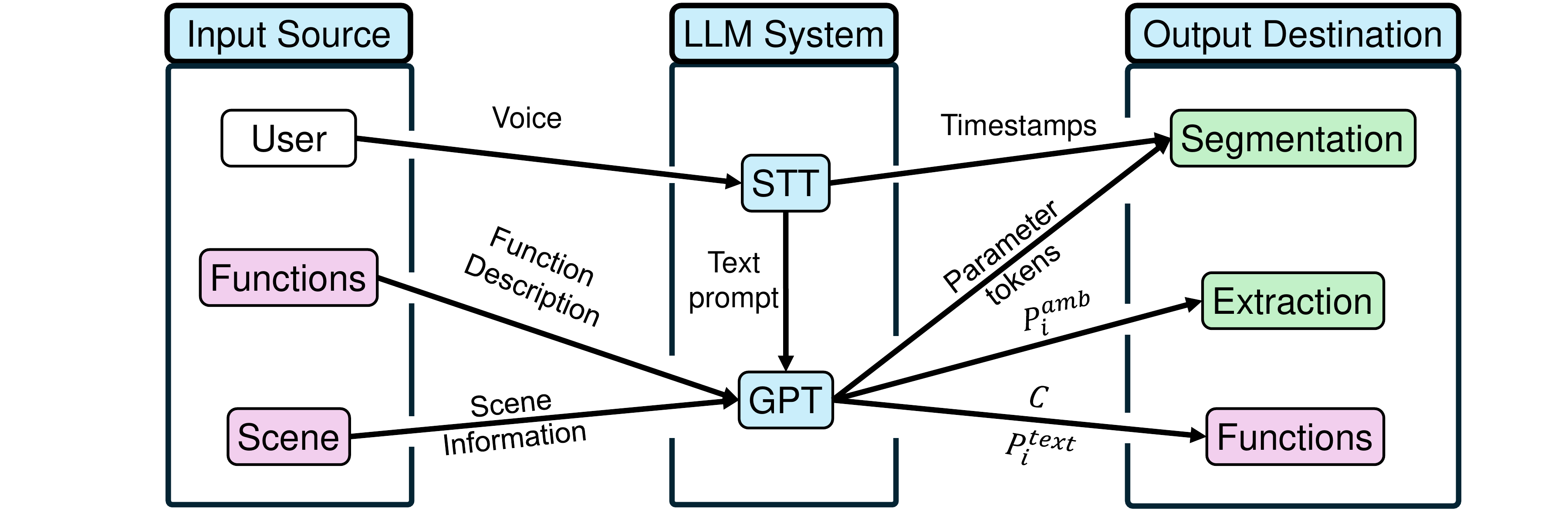}
    \caption{LLM System consists of a speech-to-text (STT) module and a GPT module. It processes user voice input and data from the XR environment. It then outputs timestamps, parameter values, and tokens to the gesture processor and the XR environment for further processing.
}
\Description{A three‑column flow diagram. The left column, “Input Source,” lists three inputs: a User box feeding “Voice” to the central STT module, a Functions box labeled “Function Description” feeding to the central GPT module, and a Scene box labeled “Scene Information” also feeding to GPT. In the middle column, “LLM System,” STT converts voice to a text prompt and emits timestamps; GPT takes the text prompt plus function descriptions and scene information and produces three outputs: ambiguous parameter tokens, resolved text parameters, and a sequence of function calls (C). The right column, “Output Destination,” shows the timestamps going to “Segmentation,” the ambiguous tokens going to “Extraction,” and the function calls going to “Functions.”}
    \label{fig:workflow_LLM}
\end{figure}

The LLM system in the workflow is composed of two key modules, as shown in \autoref{fig:workflow_LLM}: the speech-to-text (STT) module and the GPT module. 

The STT module serves as a pre-processor for users' speech. Specifically, it converts users'  speech into text and records the timestamp for each word. After the processing, the converted text is forwarded to the GPT module for further analysis, and the timestamps of words are forwarded to the \textit{Gesture Processor} for receiving gesture inputs.

Given the converted text from the STT module, along with scene information and function descriptions-including each function's purposes and \textit{function parameters} (\autoref{fig:workflow_LLM}) -the GPT module interprets the converted text to determine appropriate functions and corresponding parameter values so that the XR environment can be changed to users' intended state. 
In detail, the GPT module consists of three steps:
(1) Convert the user prompt into a sequence of function calls.
(2) For each function call, attempt to identify the values of the parameters in the prompt. 
(3) Forwards \textit{ambiguous parameters} $p_i^{amb}$ to \textit{Gesture Processor} for further calculation.
The system prompt of the GPT module is included in Appendix \ref{app:1}.

Now we illustrate how the GPT module accomplishes the above steps. Based on the user's prompt, the GPT module first determines which function calls are necessary to control the XR environment's state. For complex prompts that require multiple function calls, the GPT module breaks the prompt into sub-prompts, allowing each to be resolved by an individual function call. Next, the GPT module traverses users' prompts to extract the appropriate values for the function parameters $P_{i}^{text}$. 
The types of \(P_{i}^{text}\) extend beyond the types of spatial parameters. Any parameter value that can be determined from the user's speech can be considered as \(P_{i}^{text}\), such as colors, numeric values, and basic shapes.

If a parameter’s value cannot be identified due to ambiguity in the textual description, it is marked as \textit{ambiguous parameters} $P_{i}^{amb}$, and the parameters have the following constraints:
\begin{equation}
P_{i}=P_{i}^{text} \cup P_{i}^{amb}
\end{equation}
and
\begin{equation}
P_{i}^{text} \cap P_{i}^{amb}= \emptyset
\end{equation}
It means that the $P_{i}$ is composed of $P_{i}^{text}$ and $P_{i}^{amb}$, while there is no overlap between $P_{i}^{text}$ and $P_{i}^{amb}$.

Then, the $P_{i}^{amb}$ and corresponding \textit{parameter tokens} will be passed to the \textit{Gesture Processor} for further processing.



\subsection{\textit{Gesture Processor}}

\begin{figure}[h]
    \centering
    \includegraphics[width=\linewidth]{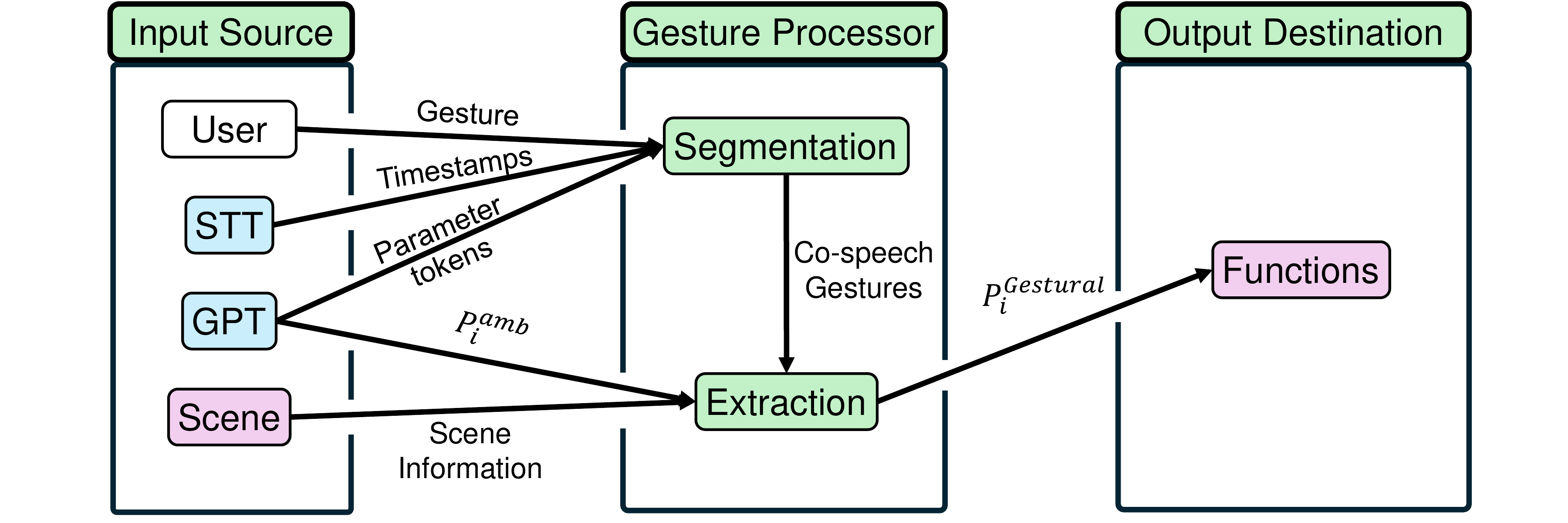}
    \caption{Gesture Processor comprised of a segmentation module and an extraction module. It handles input from the user's gestures, parameter data from the LLM system, and scene information from the XR environment. It then generates gesture parameters for integration within the XR environment.}
    \Description{Flow diagram showing the gesture processing pipeline. The left column, “Input Source,” includes four inputs: the User (sending raw Gesture data), the STT module (sending Timestamps), the GPT module (sending ambiguous parameter tokens), and the Scene (sending Scene Information). Arrows from User, STT, and GPT converge on the “Segmentation” box in the middle column labeled “Gesture Processor.” An arrow labeled “Co‑speech Gestures” leads from Segmentation to the “Extraction” box. An arrow from Scene Information also enters Extraction. Finally, Extraction outputs a parameter labeled p_gestural_i to the “Functions” box in the right column labeled “Output Destination.”}
    \label{fig:workflow_ges}
\end{figure}

We designed a \textit{Gesture Processor} to retrieve the spatial-temporal parameters. Specifically, given $P_{i}^{amb}$, \textit{parameter tokens}, and timestamps of each word from the LLM system, the \textit{Gesture Processor} analyzes gestural input and calculates the values for the \textit{ambiguous parameters} $P_{i}^{amb}$. 

The \textit{Gesture Processor} consists of two components: segmentation and extraction. As shown in \autoref{fig:workflow_ges}, in the segmentation step, the \textit{Gesture Processor} first retrieves user gestures using timestamps of \textit{parameter tokens}. For example, in \autoref{fig:teaser} (b),  the user says 'enlarge the painting to this large'. \textit{Gesture Processor} segments users' gestures when user says \textit{parameter tokens} 'the painting' and 'this large'. In the subsequent extraction step, based on the parameter type, the \textit{Gesture Processor} uses the method described in Section 3.2 to calculate parameter values ($P_{i}^{Gestural}$) from the segmented gesture data. 

\subsubsection{Segmentation} People tend to semi-automatically synchronize key phrases (\textit{parameter tokens}) with the co-speech gesture \cite{holle2008neural,wagner2014gesture}. During the segmented co-speech gesture, the gesture encapsulates the majority of the meanings integral to the speech \cite{ghaleb2024co, sanchez2022gesture, kita1998movement}. Based on these findings, \oursystem~ utilizes the timing of when \textit{parameter tokens} are articulated to extract the corresponding second stage of the co-speech gesture, subsequently processing the gesture according to the parameter type. 

\begin{figure}[h]
    \centering
    \includegraphics[width=\linewidth]{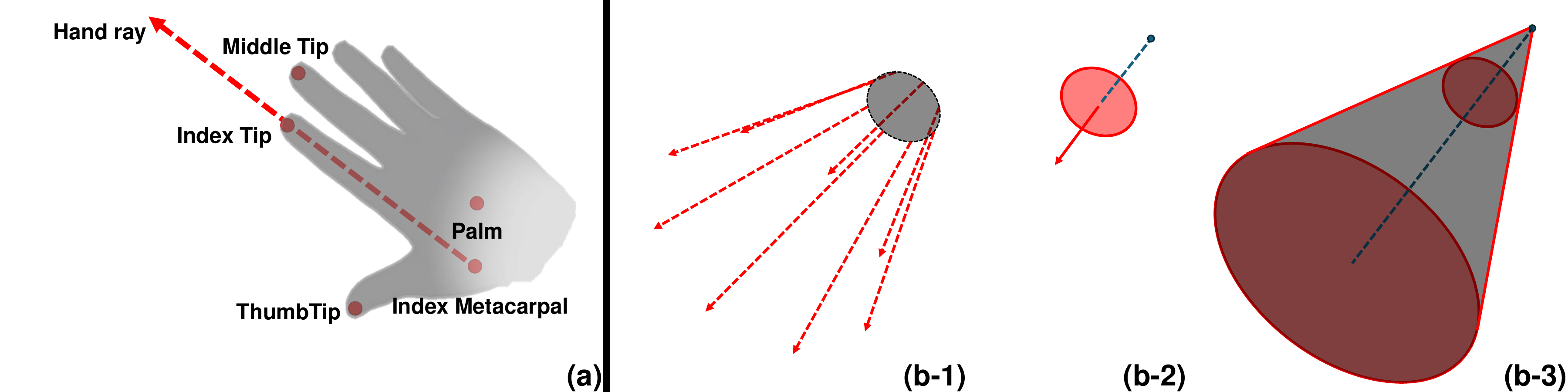}
   \caption{(a) Hand joints and hand ray used in \oursystem. (b) Process of creating a cone to select multiple object: (b-1) a set of original hand rays from the segmented gesture; (b-2) the average direction of the hand ray and the circle fit to the hand ray origins, the vertex is 30 cm from the circle; (b-3) Cone volume generated based on the vertex and circle, the height of the cone is set to 10 m.}
   \Description{A schematic in two parts. On the left (a) is an outline of a right hand showing labeled joints: palm, thumb tip, index metacarpal, index tip, and middle tip. A dashed red arrow labeled “hand ray” extends from the index metacarpal through the index tip. On the right are three panels illustrating how a cone is constructed from multiple hand rays: (b‑1) several dashed red rays from different hand positions converging toward a dark circle; (b‑2) the average direction ray in red pointing to the circle, with the circle fitted to the fingertip origins; (b‑3) a translucent red cone whose base is the fitted circle and whose vertex lies 30 cm behind that plane, with a dashed axis line showing the cone’s centerline.}
    \label{fig:Raycast}
\end{figure}

\subsubsection{Extraction}
For the gesture data, the positions and rotations of the thumb, index, middle tip joints, index metacarpal joints, and the palm joints for both hands are recorded (\autoref{fig:Raycast}(a)). The difference in palm positions between each frame and the initial frame is computed. A threshold is then applied to this difference to determine if there is hand movement within the segment.
If the segmented gesture corresponds to spatial parameters that can be expressed by the deictic gestures, the hand ray (\autoref{fig:Raycast}(a)) will be used to determine the parameter value. The hand ray originates from the index metacarpal joint and points to the index fingertip joint. The length of the ray is set to 10 m. For the position parameter, we calculate the average position of intersection points. 

For the object parameter, a cone-shaped bounding volume is generated to encompass all intersecting objects. This is accomplished using the method illustrated in \autoref{fig:Raycast}(b). From a segment of gesture data, the hand ray is computed for each frame, as shown in \autoref{fig:Raycast}(b-1). The cone's directional vector is derived by averaging these hand-ray vectors. Assuming the index tip positions describe a circle, this circle is fit using the least-squares method on a plane, with the plane's normal corresponding to the directional vector and the average palm positions lying on it. Empirically, the vertex of the cone is positioned 30 cm behind this plane, as indicated by the blue point in \autoref{fig:Raycast}(b-2). The cone's height is defined as 10 m. Using the directional vector, circle, vertex, and height parameters, the cone-shaped volume is formed, as depicted in \autoref{fig:Raycast}(b-3).

For the direction parameter, the average hand ray is calculated to be the vector for this parameter.

For the rotational parameter, if a single hand is involved, the rotational difference of the palm between the initial and the final frame is utilized. For gestures involving two hands, the line connecting the palm joints is considered, and the rotational difference of this line between the first and the last frame is used as the parameter's value.

For the size parameter in one-hand gestures, the average distance between the index fingertip joint and the thumb tip joint is utilized. If the average distance between the index tip and middle tip joints falls below a certain threshold, a ray is projected from the palm joints. Unlike the hand ray, the direction of this ray follows the palm joint's forward direction (the direction the palm faces) and terminates at the closest surface. The average length of these rays will determine this parameter value. In the case of two-hand scenarios, the mean distance between the palm joints of each hand is considered.

For the path parameter, the positions and rotations of the six joints for each hand are utilized as the parameter value.



\subsection{XR System}

\begin{figure}[h]
    \centering
    \includegraphics[width=\linewidth]{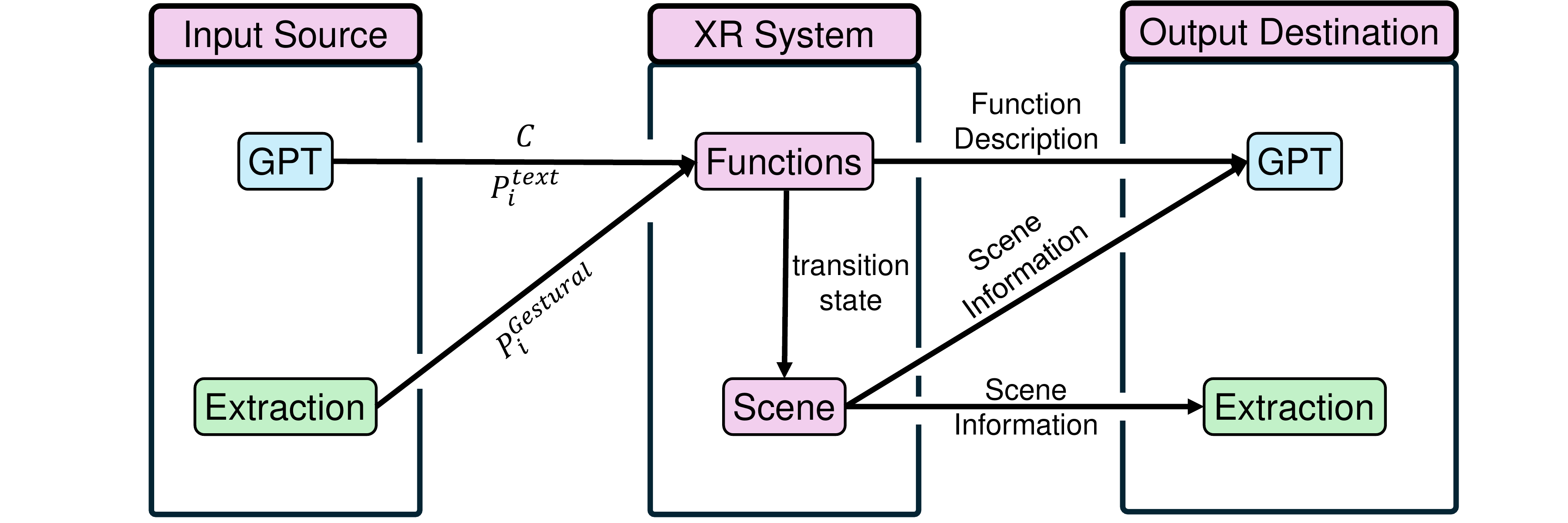}
    \caption{XR system includes available functions and scene information. It provides and obtains data to the LLM system and gesture processor, and updates the scene state correspondingly.}
    \Description{Three‑column diagram showing how function calls and scene data flow through the XR system: Left column labeled “Input Source” contains two boxes: “GPT” (blue) and “Extraction” (green). An arrow from GPT to the center “Functions” box is labeled “C, P\_text\_i”. A second arrow from Extraction to Functions is labeled “P\_gestural\_i”. Center column labeled “XR System” contains two stacked boxes: “Functions” (purple) above “Scene” (purple). From Functions there is a downward arrow labeled “transition state” into Scene. From Functions there is also a rightward arrow to the right column GPT labeled “Function Description”. From Scene two arrows go right: one to the rightmost “GPT” box labeled “Scene Information” and one to the rightmost “Extraction” box also labeled “Scene Information”. Right column labeled “Output Destination” contains “GPT” (blue) and “Extraction” (green).}
    \label{fig:workflow_xr}
\end{figure}

The \textit{XR System} manages available functions and scene information as shown in \autoref{fig:workflow_xr}.
It receives function calls $C$ and associated parameters $P_{i}^{text}$ from the LLM system, as well as $P_{i}^{Gestural}$ from the \textit{Gesture Processor}. If any parameters remain unresolved, the user will be prompted for further clarification through speech or gestures. \oursystem~ presents a text box requesting the user to provide clarification, as illustrated in \autoref{fig:Framework} failed cases. The error message details the function that \oursystem~ intends to execute, along with the parameters that are missing. When the user attempts to call functions that are not available in the XR system, a general error message will be shown to the user (``Sorry, the system is unable to do that, the system is able to do [list of functions].''). Once all parameters are confirmed, the \textit{XR system} executes the function calls to transition the scene to the user's intended state.

In the prototype system, a set of XR functions that assist fundamental XR manipulation was implemented to showcase the feasibility of \oursystem. Researchers in the future may develop specialized functions to facilitate additional interactions or opt for a system such as LLMR~\cite{de2024llmr}, which employs LLMs to generate scripts derived from the user's prompt. When an XR function is added to the XR system, it is necessary to revise the metaprompt for the GPT module within the LLM system to incorporate the function's description and parameters.

Here we summarize the XR functions that are implemented in the prototype system:

\begin{itemize}
    \item select(list of objects, color): selects objects of a certain color from a list of objects.
    \item move(object, position): move an object to a new location.
    \item rotate\_dir(object, direction): set the forward direction of the object to the target direction.
    \item rotate(object, rotation): applies the rotation to the object.
    \item resize(object, size): change the size of the object to the target size
    \item move\_path(object, path): let an object move along a trajectory back and forth.
    \item draw\_path(path, shape\_type): sketch a predefined shape given the path. The shape\_type includes straight lines, circles, and sine waves. We employ a linearized least-squares approach for circle fitting, ordinary least-squares regression for straight lines, and nonlinear least-squares optimization for sine waves.
    \item set\_color(object, color): set the color of an object to a given color.

\end{itemize}

The scene information sent to the GPT module is formatted in JSON and contains the name, position (meters), rotation (degrees), and scale (meters) of every object present in the scene. Each parameter's numerical value is rounded to three decimal places. An example can be found in Appendix \ref{app:3}.

\subsection{Implementation}
We developed a prototype VR system that implements the proposed framework in Sec \ref{sec:s3}. 
The application is developed on the Unity3D platform (2022.3.20f1) \cite{unity} and runs on the Meta Quest Pro headset \cite{Oculus}.
The user's voice input is converted into text using the Azure AI speech service \cite{azure}.
For the LLM system, we used GPT-4o model \cite{gpt-4o}~ to process the given text.
Skeletal gesture data are acquired using native sensors from the Meta Quest Pro headset and the Meta XR SDK \cite{meta-sdk}.

\section{User Study}

We conducted a three-session IRB-approved user study to evaluate: (1) the accuracy of combining natural language and gesture input using \oursystem~,  (2) the difference between \oursystem, gesture-only system, and voice-only system, and (3) the usage of \oursystem~  in interior design scenario.
The entire study lasted approximately 1.5 hours, and each participant received a $\$20$ e-gift card. The study was conducted in a $5$m by $5$m indoor area and was screen-recorded for post-analysis. 
Upon arrival, participants were given an introduction to the system, including a detailed explanation of the workflow and user interface. 
After three sessions, each participant completed a 5-point Likert-scale questionnaire and the standard System Usability Scale (SUS). 
We concluded the study with a conversation-style interview to gather subjective feedback on the system.

\subsection{Participants}

We recruited 12 participants (ten males and two females), ranging in age from 18 to 26.
Of these, 1 is an experienced developer of head-mounted AR/VR applications, 8 had some experience with head-mounted AR/VR applications, while the remaining three had only heard of AR and VR.
Four participants use ChatGPT often, and eight participants use ChatGPT almost every day. Additionally, two participants had no experience with Prompt Engineering, five participants had heard of this concept, two participants had tried it before, and three participants almost always uses prompt engineering in communicating with ChatGPT. None of the participants had used our system prior to the study.

\subsection{Session 1}
\begin{figure}[h]
    \centering
    \includegraphics[width=\linewidth]{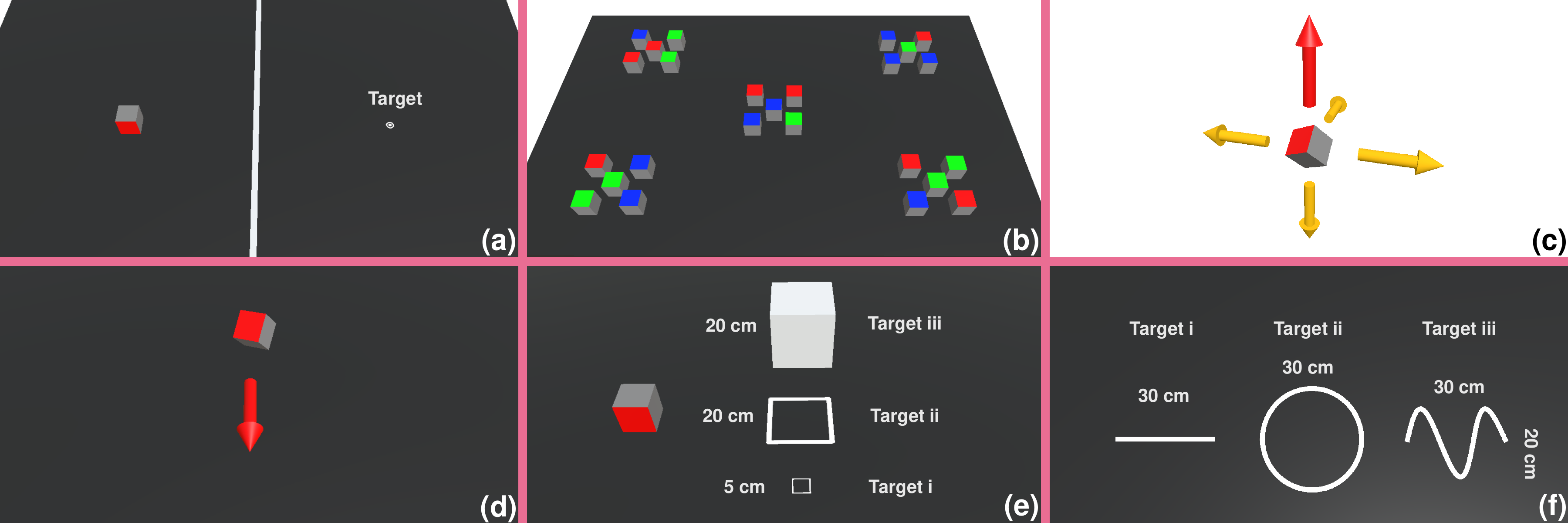}
    \caption{User study session 1 setup: (a) position task; (b) object task; (c) direction task; (d) rotation task; (e) size task; (f) path task.}
    \Description{Six panels labeled (a) through (f) showing the six spatial‑parameter tasks in Session 1. Panel (a) shows a single red‑topped cube on the left of a dividing line and a small white dot labeled “Target” on the right. Panel (b) shows 25 small cubes in five clusters of red, green, and blue on a table surface. Panel (c) shows a central cube with yellow arrows pointing outward in all horizontal directions and a red arrow pointing upward. Panel (d) shows a red‑topped cube hovering above a down‑pointing red arrow. Panel (e) shows three target cubes and a wireframe: a large white cube labeled “Target III” with 20 cm markers, a medium cube outline labeled “Target II” with a 20 cm marker around a red cube, and a small red cube labeled “Target I” with a 5 cm marker. Panel (f) shows three path shapes labeled “Target I,” “Target II,” and “Target III”: a 30 cm horizontal line, a 30 cm‑diameter circle, and a sine wave 30 cm long and 20 cm tall.}
    \label{fig:UserStudy_1_setup}
\end{figure}

\subsubsection{Procedure}

In this session, we evaluated the system's performance in analyzing both speech and gesture data. In alignment with earlier studies on fundamental object manipulation within virtual environments \cite{chae2018wall, mendes2016benefits}, we designed six tasks to examine each spatial parameter independently: 
\begin{enumerate}
    \item \textbf{Position}: move a cube on the left side of the table to the target  (\autoref{fig:UserStudy_1_setup}(a)). The initial position of the cube is randomized in the left region.
    \item \textbf{Object}: select cubes of a certain color from piles of cubes (\autoref{fig:UserStudy_1_setup}(b)). A total of 25 colored cubes, 9 red, 8 green, and 8 blue on the table. The target color and the position of the cubes are randomized for each trial. We count the number of correct cubes selected (TP), incorrect cubes selected (FP), not selected correct cubes (TN), and not selected incorrect cubes negative selection (FN). 
    \item \textbf{Direction}: set the front face (marked in red) of the cube to one of the five directions (\autoref{fig:UserStudy_1_setup}(c)). The initial rotation of the cube is randomized in 360$^\circ$ on the x, y, z axis. The target direction is highlighted in red. The order of the target direction is randomized for each participant. 
    \item \textbf{Rotation}: rotate a cube so that the front face is facing the participant (\autoref{fig:UserStudy_1_setup}(d)). The initial rotation of the cube is randomized in 180$^\circ$ on the x, y, z axis, and the red side is always visible to the participant. Half of the participants are asked to use one hand, and the other half to use two hands to rotate the object.
    \item \textbf{Size}: resize the cube to the size of the target (\autoref{fig:UserStudy_1_setup}(e)). The initial length of the cube is set to 10 cm. 
    One-third of the participants each are asked to use (i) one hand (target length 5 cm), (ii) two hands (target length 20 cm), and (iii) use the table as a reference (target length 20 cm) to resize the object.
    \item \textbf{Path}: create a path base on the reference path (\autoref{fig:UserStudy_1_setup}(f)).  \added{The reference path is shown on the table. The participants were asked to draw the path mid-air using the index finger of their dominant hand while giving the speech command.}
    One-third of the participants each are asked to define (i) a straight line (target length 30 cm), (ii) a circle (target diameter 30 cm), and (iii) a sine wave (target length 30 cm, amplitude 20 cm, period of 1.5).
\end{enumerate}
In order to evaluate one spatial parameter per task, the cube object will be automatically chosen for tasks aside from the object task. Additionally, following prior work \cite{wang2021gesturar,de2024llmr}, the following evaluation metrics for failure cases are also used: (1) whether the LLM system could identify $P_i^{amb}$ and the \textit{parameter tokens} (type 1 error) and (2) whether the gesture data were correctly analyzed based on the associated parameters and phrases (type 2 error). A failed attempt is characterized by the authors being able to comprehend the participant's command through speech and gesture, recognizing it as appropriate for the task, yet \oursystem~ does not generate an accurate outcome. Participants were instructed to execute each task five times. The successful trials were utilized to assess the parameter metrics, while unsuccessful attempts were included for error analysis. Completion time refers to the interval between the moment the participant initiates the command and when the system finalizes the execution of the function calls. The Mann-Whitney U test is used to compare the results.

Participants were asked to treat the target as a mental reference, and they should not explicitly mention it in their commands (i.e., they should not say "move the cube to the target" without any gesture).
The LLM system does not have information about the target.
We instructed the participants to interact with the system as if they were communicating with a human and to request the system to adjust the cube to the target state. The participant is informed of the task type before each trial. trail
No additional text or voice instructions were provided to minimize influence on their commands.

\subsubsection{Results and Discussion}

\begin{figure}[h]
    \centering
    \includegraphics[width=\linewidth]{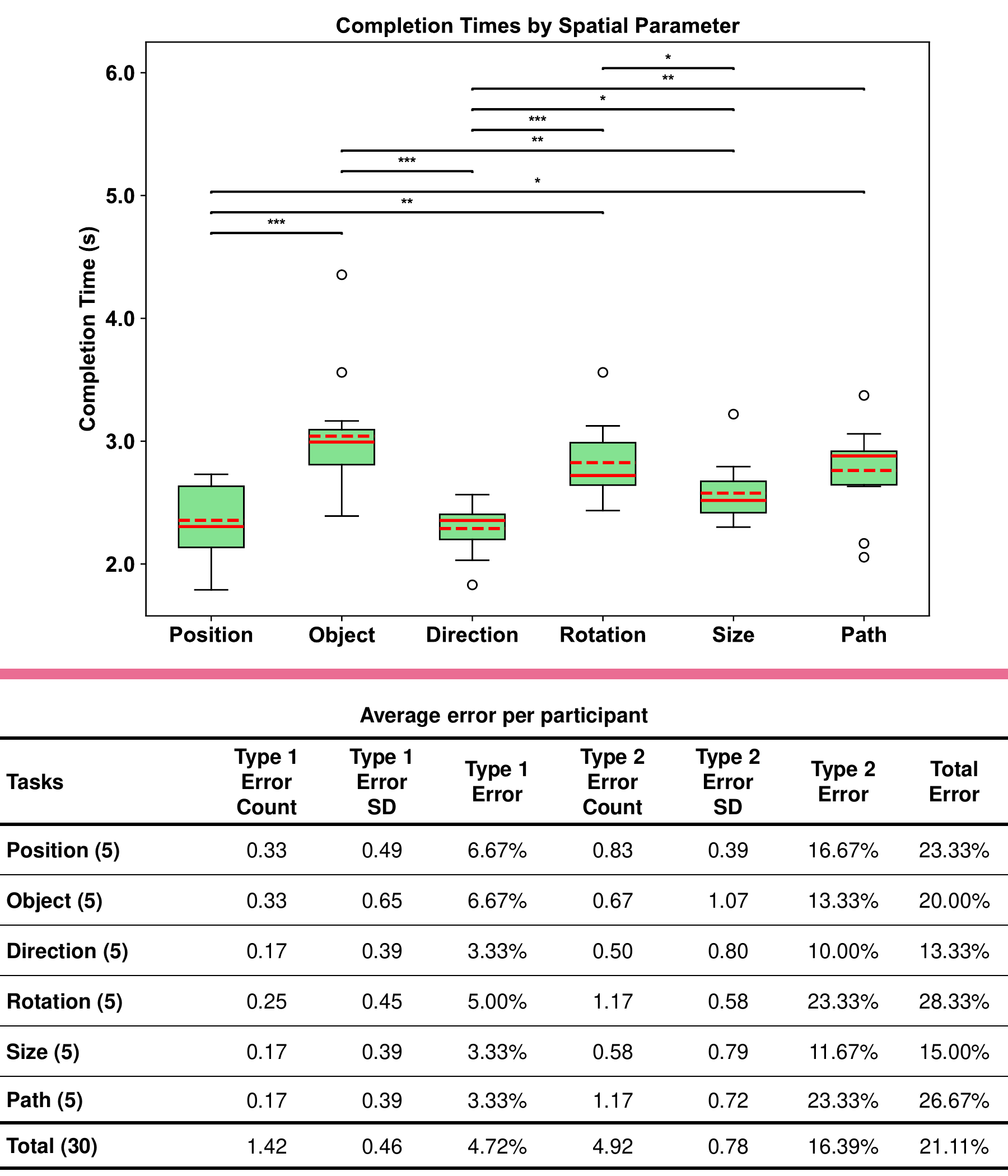}
    \caption{User study session 1 result. Top: completion times by spatial parameter, ``*'' indicates p-value <0.05, ``**'' indicates p-value < 0.01, ``***'' indicates p-value < 0.001. The red solid line represents the median, and the red dashed line represents the mean. Bottom: average error per participant.}
    \Description{Top panel shows a box‑and‑whisker plot titled “Completion Times by Spatial Parameter (s)” with six categories: Position, Object, Direction, Rotation, Size, and Path. Each category has a green box indicating the interquartile range, a red line for the median, whiskers for the 1.5× IQR range, and individual outlier circles. Direction has the lowest median (\~2.3 s), Position has the highest (\~2.7 s), and other parameters fall between \~2.5 s and 3.2 s. Multiple horizontal lines with asterisks above the boxes denote pairwise significance: three asterisks for p<0.001, two for p<0.01, one for p<0.05. Bottom panel is a table titled “Average error per participant” listing six tasks (Position, Object, Direction, Rotation, Size, Path) with counts and standard deviations for Type 1 and Type 2 errors, as well as percentages, plus a Total error row. }
    \label{fig:UserStudy_1_result}
\end{figure}


All users completed the tasks with an average completion time per command of 3.28 seconds (SD = 0.85) and an average total error rate of \(21.11\%\). The quantitative comparison results of completion time for each task are shown in \autoref{fig:UserStudy_1_result} top. We observe that 
the average completion time for commands involving parameters ``Position'' (AVG = 2.71, SD = 0.60), ``Direction'' (AVG = 2.58, SD = 0.44), ``Size'' (AVG = 3.15, SD = 0.50) are lower than ``Object'' (AVG = 4.08, SD = 1.03), ``Rotation'' (AVG = 3.65, SD = 0.63), ``Path'' (AVG = 3.52, SD = 0.72). Of the system errors, shown in \autoref{fig:UserStudy_1_result} bottom, the type 1 error rate out of all attempts is 4.72\% and the type 2 error rate is 16.39\%. The highest total error rate occurred in the rotation task (28.33\%), followed by the path task (26.67\%), position task (23.33\%), object task (20.00\%), size task (15.00\%), and direction task (13.33\%). The gestures used in the object, rotation, and path task are dynamic, implying that even a minor alteration in gesture can lead to significant changes in the outcome. Thus, tasks with dynamic gestures require more time.  At the same time, accurately segmenting the portion of the gesture when it occurs is crucial in analyzing the dynamic gesture data, as it contributes to an elevated error rate in these tasks.

\begin{table}
  \caption{Summary of quantitative measurements for different spatial parameters. 
  Position represents the distance between the object and the target's placement. 
  Direction captures the angular difference between the object and the target. 
  Rotation indicates the least rotation required to align the object with the target. 
  Size measures the percentage difference in size between the object and the target. 
  Path evaluates the similarity between the user's path and the intended path using DTW-based similarity.}
  \Description{A dark tabletop scene showing two cube arrangements. On the left, a small cluster of gray cubes with one cube’s face colored red is labeled “Object” by a pink arrow. On the right, a ring of gray cubes with a few red faces is arranged in a circle 0.6 m across and labeled “Target” by a pink arrow.}
  \label{tab:quantitative}

\begin{tabular}{lcc}
\toprule
\textbf{Parameter} & \textbf{Mean} & \textbf{STD} \\
\midrule
Position (m)         & 0.0387 & 0.0161 \\
Direction ($^{\circ}$)  & 2.37   & 2.30 \\
Rotation ($^{\circ}$)   & 12.52  & 8.93 \\
Size (\%)             & 16.84  & 12.36 \\
Path (\%)             & 91.47  & 8.23 \\
\toprule
\multicolumn{3}{l}{\textbf{Object Selection}} \\
\midrule
Precision (\%)       & 100.00  &\\
Recall (\%)         &96.56    & \\
\bottomrule
\end{tabular}
\Description{A table of system accuracy metrics with two sections. The first section lists spatial parameters and their mean error and standard deviation: Position (m): mean 0.0387, STD 0.0161; Direction (°): mean 2.37, STD 2.30; Rotation (°): mean 12.52, STD 8.93; Size (\%): mean 16.84, STD 12.36; Path (\%): mean 91.47, STD 8.23. The second section, titled “Object Selection,” reports precision 100.00\% and recall 96.56\%.}
\end{table}

\begin{table}
  \caption{Quantitative measurements for rotation, size, and path tasks under different conditions. Rotation (i): one hand; Rotation (ii): two-hand; Size (1): one hand; Size (ii): two-hand; Size (iii): refer to the table; Path (i): straight line; Path (ii): circle; Path (iii): sine wave}
  \label{tab:quantitative_detail}
\begin{tabular}{lcc}
\toprule
             & Mean         & STD         \\
\midrule
Rotation (i)  & 17.89$^{\circ}$  & 8.17$^{\circ}$  \\
Rotation (ii) & 10.82$^{\circ}$  & 8.22$^{\circ}$  \\
Size (i)      & 23.41\%      & 11.39\%      \\
Size (ii)     & 12.54\%      & 11.03\%      \\
Size (iii)    & 18.18\%      & 13.28\%      \\
Path (i)      & 95.25\%      & 2.34\%       \\
Path (ii)     & 90.67\%      & 3.45\%       \\
Path (iii)    & 88.50\%      & 7.37\%       \\
\bottomrule
\end{tabular}
\Description{A table showing mean and standard deviation for rotation, size, and path across different conditions: Rotation (i): mean 17.89° ± 8.17°; Rotation (ii): 10.82° ± 8.22°; Size (i): 23.41\% ± 11.39\%; Size (ii): 12.54\% ± 11.03\%; Size (iii): 18.18\% ± 13.28\%; Path (i): 95.25\% ± 2.34\%; Path (ii): 90.67\% ± 3.45\%; Path (iii): 88.50\% ± 7.37\%.}
\end{table}

For successful trials, we assess the accuracy of the results for individual tasks (\autoref{tab:quantitative}). The position metric is determined by comparing the object's final position with its target (AVG = 3.87 cm, SD = 1.61 cm). The system exhibits a good understanding of the intents involved in the position task. For object metrics, we evaluate precision (\(\frac{TP}{TP+FP}\)) and recall (\(\frac{TP}{TP+FN}\)). The average precision achieved is 100\%, indicating that all selected cubes are of the correct colors, demonstrating that \oursystem~ can comprehend object attributes from speech. The average recall is 96.56\%, suggesting that the selection of objects by cone misses some objects outside of it, likely due to an incomplete gesture segment. For direction and rotation, the same degree difference metric is employed. We calculate the least amount of rotation needed for the object to rotate to the target rotation. Compared with the direction task (AVG = 2.37$^\circ$, SD = 2.30$^\circ$), the rotational difference is significantly higher for the rotation task (AVG = 12.52$^\circ$, SD = 8.93$^\circ$, $p$ = 0.0025). This discrepancy could be attributed to the fact that the rotation task requires dynamic gestures, whereas the meaningful segment of the gesture in the direction task is more static. Furthermore, since the cube remains stationary during the command, it may leave the user uncertain about how much hand rotation is necessary. For the size parameter, we calculated the percentage size difference to assess its accuracy, i.e., \(\frac{||Object Length-Target Length||}{Target Length}\). For the path parameter, dynamic time wrapping-based similarity was used as a metric. The average similarity is 91.47\% and the standard deviation is 8.23\%.

\autoref{tab:quantitative_detail} provides a detailed examination of system accuracy for rotation, size, and path tasks under various conditions. Performance with two hands (AVG = 10.82$^\circ$, SD = 8.22$^\circ$) in the rotation task is more precise than with one hand (AVG = 17.89$^\circ$, SD = 8.17$^\circ$), though not significantly. For the size task, two-hand usage results in a smaller size percentage difference (AVG = 23.41\%, SD = 11.39\%) compared to one-hand (AVG = 12.54\%, SD = 11.03\%, $p$ = 0.044) and the reference (AVG = 18.18\%, SD = 13.28\%, $p$ = 0.035). In the path task, drawing a line leads to higher similarity (AVG = 95.25\%, SD = 2.34\%) than drawing circles (AVG = 90.67\%, SD = 3.45\%, $p$ = 0.0031) or sine waves (AVG = 88.50\%, SD = 7.37\%, $p$ = 0.004).

\subsection{Session 2}

\begin{figure}[h]
    \centering
    \includegraphics[width=\linewidth]{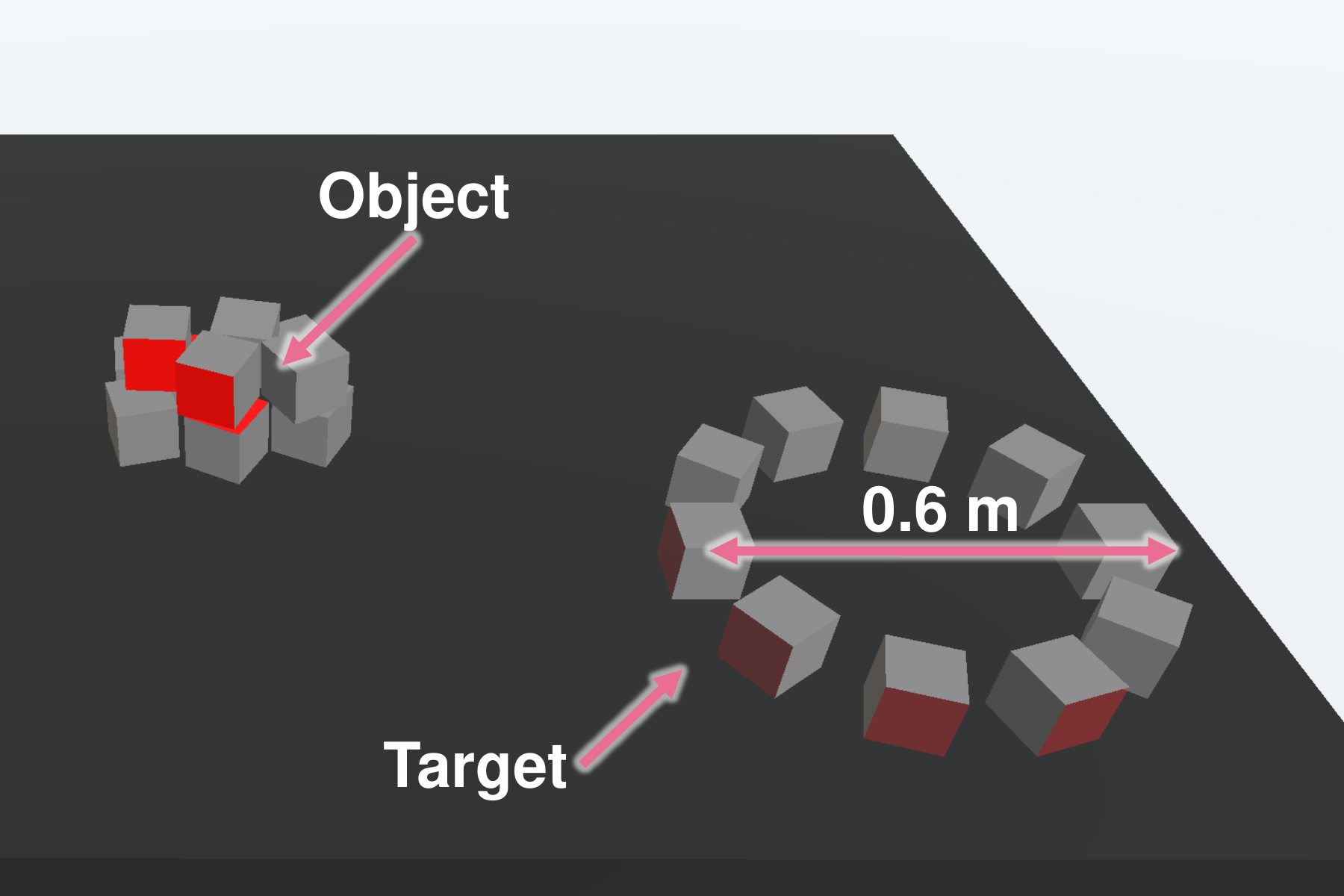}
      \caption{\added{User study session 2 setup}}
    \label{fig:UserStudy_2_setup}
\end{figure}

\subsubsection{Procedure}

In this session, we evaluated the difference between \oursystem~ and two baseline systems.
Guided by \cite{ledo2018evaluation,de2024llmr}, we designed an A/B comparison study to compare user satisfaction when using \oursystem, gesture-only system, and voice-only system to complete the same spatial task, which is to manipulate the 10 red cubes on the table so that they match the target state. The setup for session 2 can be found in \autoref{fig:UserStudy_1_setup}.
The task is structured to encompass all six spatial parameters and is executable with all three systems.
This task incorporates all six spatial parameters as previously indicated: (1) the user is required to select an \textbf{object}, (2) generate a circular \textbf{path}, (3) define the \textbf{position} of this path, (4) determine the \textbf{size} (diameter) of the circle, (5) as well as establish the \textbf{rotation} or \textbf{direction} of each cube.

For the gesture-only system, only the manipulative gestures are enabled.
The user is restricted from using predefined Meta XR SDK hand gestures to grab and release an object. Once the object is grabbed, it will follow the position and rotation of the user's hand. While with \oursystem, none of the predefined gestures are enabled. \oursystem~ only takes speech and raw gesture data as input.
For the voice-only system, we remove the gesture processor from \oursystem. The metaprompt, shown in Appendix \ref{app:2}, is modified to instruct the GPT agent to find the values of all parameters solely on the basis of the prompt and scene information given. 
We asked the user to treat the target as a mental reference and gave them the same instructions as in session 1.
The time limit for using each system was restricted to 10 minutes.
We recorded the total interaction time with the system.
For the gesture-only system, the interaction time is defined as the time between the first manipulation of the object and the last manipulation.
For the other two systems, the interaction time is defined as the time between the first command and the last command.
The order of use of \oursystem~ and the baseline systems was randomized for each user. At the start of the task, the user was informed of which system they were using. The user was directed to arrange the cubes in a configuration as similar as possible to the target. They were allowed to stop whenever they believed they had achieved their best effort in organizing the cubes. We applied Mann-Whitney U tests to compare the results from the questionnaire.

\subsubsection{Results and Discussion}

\begin{figure}[h]
    \centering
    \includegraphics[width=\linewidth]{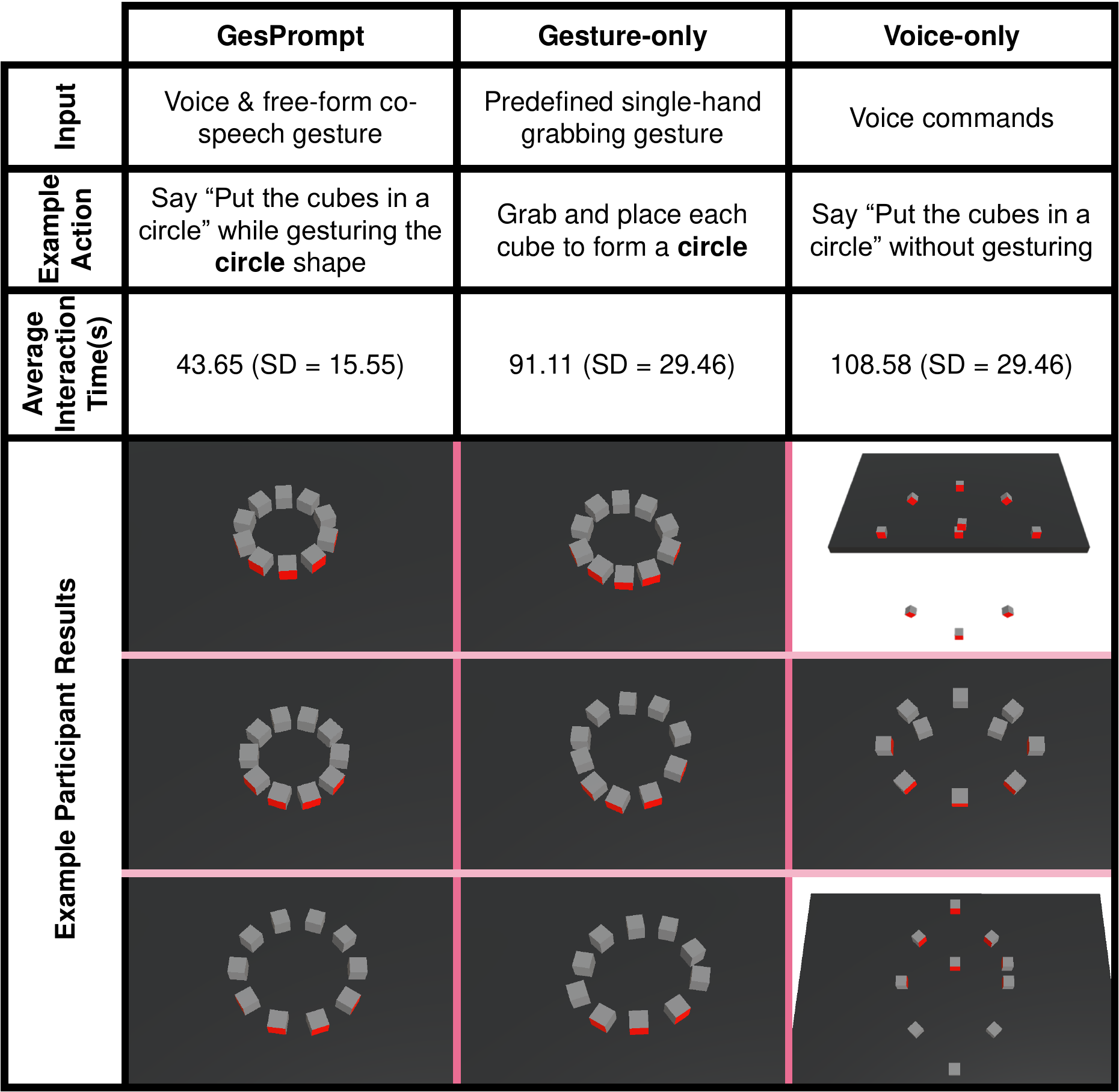}
    \caption{\added{Session 2 conditions and example participant results.}}
    \Description{A table comparing three interaction systems—GesPrompt, Gesture‑only, and Voice‑only—across actions, average interaction time, and example participant results.  Under “Actions,” GesPrompt is labeled “Voice + co‑speech gesture commands,”  Gesture‑only “Only single‑hand grabbing gesture,” and Voice‑only “Only voice commands.” Under “Average Interaction Time,” GesPrompt shows 43.65 s (SD 15.55), Gesture‑only 91.11 s (SD 29.46), Voice‑only 108.58 s (SD 29.46).   Below, three rows of example results illustrate how participants arranged red‑highlighted cubes into a circle: GesPrompt column shows three circles of gray cubes with the red target cubes accurately placed each time.   Gesture‑only column shows three circles where some red cubes are slightly misaligned.   Voice‑only column shows three attempts where cubes are scattered on the floor or misarranged in a rough circle.}
    \label{fig:UserStudy_2_table}
\end{figure}

\begin{figure}[h]
    \centering
    \includegraphics[width=\linewidth]{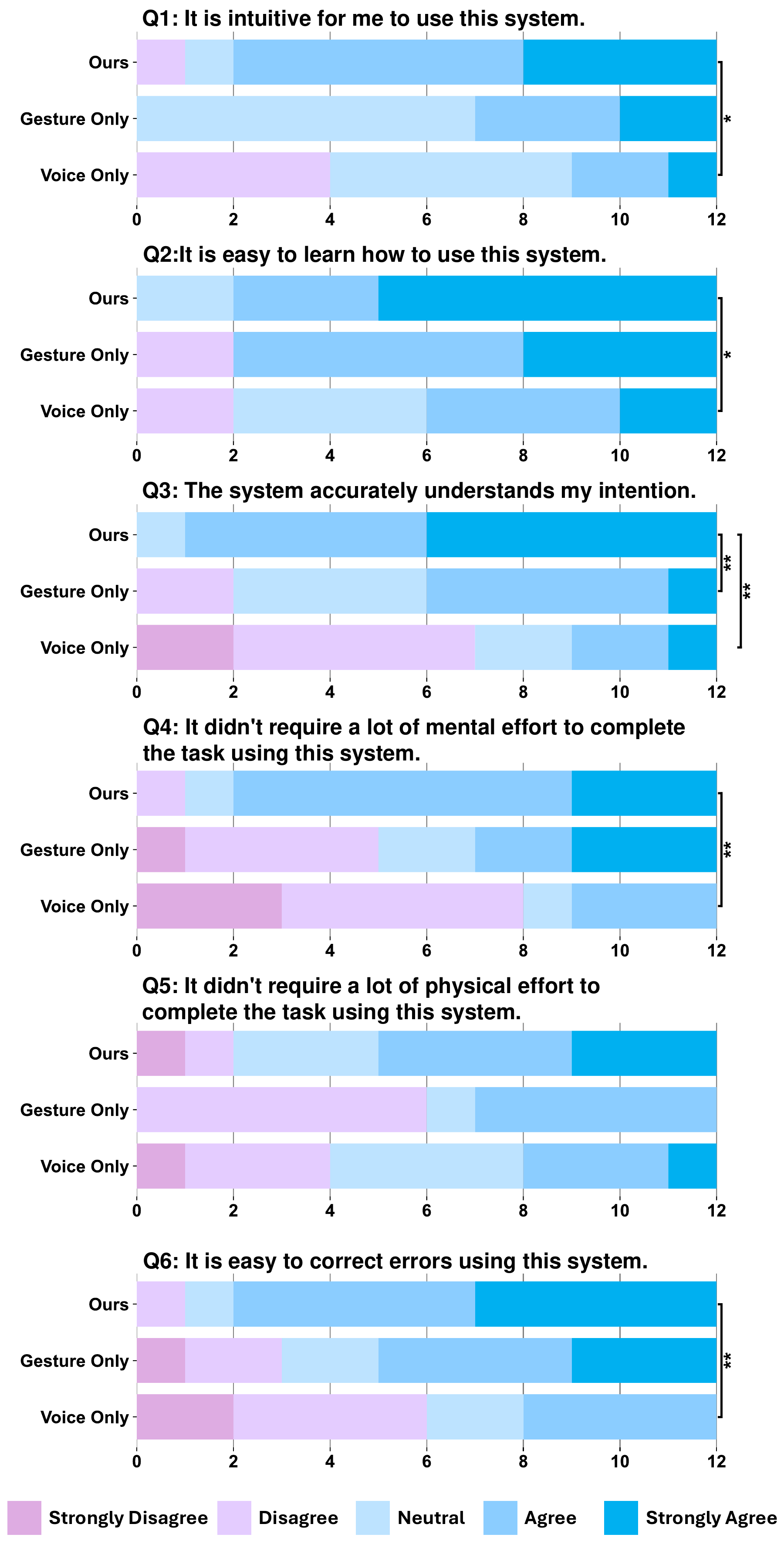}
    \caption{Session 2 Likert-type questionnaire results; ``*'' indicates p-value <0.05, ``**'' indicates p-value < 0.01.}
    \Description{Six stacked horizontal bar charts labeled Q1 through Q6, each comparing three systems (“Ours,” “Gesture Only,” and “Voice Only”) across a 12‑point Likert scale from Strongly Disagree to Strongly Agree. Bars are colored from purple (Strongly Disagree) through light purple (Disagree), light blue (Neutral), medium blue (Agree), to dark blue (Strongly Agree).  Q1 (“It is intuitive for me to use this system.”) shows “Ours” almost entirely in Agree and Strongly Agree, “Gesture Only” mostly Neutral through Strongly Agree, and “Voice Only” from Disagree to Agree, with a significance bracket between “Ours” and “Voice Only” (p < .05). Q2 (“It is easy to learn how to use this system.”) shows “Ours” entirely in Neutral, Agree, and Strongly Agree; “Gesture Only” in Disagree through Strongly Agree; “Voice Only” mostly Disagree and Neutral; with a significance bracket between “Ours” and “Gesture Only” (p < .05). Q3 (“The system accurately understands my intention.”) shows “Ours” heavily in Agree and Strongly Agree, “Gesture Only” spanning Disagree to Strongly Agree, and “Voice Only” mostly Strongly Disagree through Agree, with two significance brackets between “Ours” vs. “Gesture Only” and “Ours” vs. “Voice Only” (p < .01). Q4 (“It didn’t require a lot of mental effort to complete the task.”) shows “Ours” in Neutral to Strongly Agree, “Gesture Only” from Strongly Disagree through Strongly Agree, and “Voice Only” mainly Strongly Disagree through Neutral, with a significance bracket between “Ours” and “Voice Only” (p < .01). Q5 (“It didn’t require a lot of physical effort to complete the task.”) shows all three systems spread from Strongly Disagree through Agree without significance annotations. Q6 (“It is easy to correct errors using this system.”) shows “Ours” mostly Agree and Strongly Agree, “Gesture Only” spanning Disagree to Strongly Agree, and “Voice Only” from Disagree through Agree, with a significance bracket between “Ours” and “Gesture Only” (p < .01).}
    \label{fig:UserStudy_2}
\end{figure}


The 12 users completed the task using the three systems within the time limit. 
The Likert-type results collected from this session are shown in \autoref{fig:UserStudy_2}.
The average interaction time with each system is: gesture-only system 91.11 seconds (SD = 23.00); voice-only system 108.58 seconds (SD = 29.46); and our system 43.65 seconds (SD = 15.55). 
The SUS results show that \oursystem~ has good usability, with a mean score of 86 out of 100. Moreover, the gesture-only system obtained a SUS score of 74, while the voice-only system received a score of 59.


Overall, users found it intuitive (Q1: AVG = 4.08, SD = 0.25) and easy to learn how to use our system (Q2: AVG = 4.42, SD = 0.21). ``I would use gesture plus speech to communicate with others in my daily life.  (P5)''. In comparison to the voice-only system (Q1: AVG = 3.00, SD = 0.26; Q2: AVG = 3.5, SD = 0.28), \oursystem~ is more accessible as users are not required to construct their prompts. Our system (Q3: AVG = 4.14, SD = 0.86) demonstrates intent understanding ability higher than the gesture-only system (Q3: AVG = 3.42, SD = 0.25) and voice-only system (Q3: AVG = 2.58, SD = 0.0.34). \textit{``I did not expect that your system can actually understand what I mean by saying `that'. (P2)''}. In the gesture-only system, some users (P4, P7) with less HMD AR/VR experience report that the hand gesture for grabbing and releasing objects feels unnatural, causing frequent failures to grab their intended object. In the voice-only system, some users felt frustrated when the cubes did not line up in a circle, even after they specifically told the system to create 10 positions in a circular layout and move the cubes there. It did not require much mental effort to use our system (Q4: AVG = 4.00, SD = 0.23) compared to the voice-only system (Q4: AVG = 2.33, SD = 0.32). \textit{``I don't need to calculate or refer to the size of the table to change the diameter of the circle (with our system). I can just do a gesture. While in the voice-only system, I have to think about them and explicitly tell the GPT. (P9)''}. Regarding the physical effort, the users find \oursystem~(Q5 : AVG = 3.58, SD = 0.34) demands less than the other two systems (Gesture-only (Q5 : AVG = 2.92, SD = 0.27); Voice-only (Q5 : AVG = 3.00, SD = 0.31)). \textit{``It is frustrating to manually organize the 10 cubes in a circle (with gesture-only system). When I add a new cube to the circle, I also have to adjust the positions of other cubes to make it look like a circle. (P10)''}. Most users agree that it is easier to correct errors (Q6 : AVG = 4.17, SD = 0.25) using our system in comparison to the voice-only system (Q6 : AVG = 2.66, SD = 0.32). \textit{``I just don't know what I should say for the voice-only system to understand me. (P4)''} Yet, one user said that \textit{``I would like to know that the system understands which object I am talking about before I can give the following command. (P8)''}. We will further discuss this concern in Section \ref{sec:f1}.

Although the interaction time for the gesture-only system is close to that of the voice-only system, the resulting cube configuration using the gesture-only system did not match the target cubes. 
After placing the cubes on the edge of a rough circle, most of the users requested to skip the task and said that they were satisfied with the result. 
The longer interaction time of the gesture-only system compared to our system is due to that the user has to try multiple times to rotate the cube to a suitable orientation and space the cubes evenly along the circle.
For the voice-only system, we observed that users gave simple prompts like \textit{"put the cubes in a circle"},  expecting the LLM agent to put the cubes in a spatially reasonable position (i.e., on the table). This observation agrees with Manesh et al. \cite{aghel2024people}. Although the LLM agent has all scene information, the outcome of executing this command is that the cubes are on the floor rather than the table, and the form resembles an ellipse rather than a circular shape (\autoref{fig:UserStudy_2_table}).
As a result, the user needed to experiment and incrementally add spatial constraints to the prompt, increasing the interaction time with the voice-only system compared to \oursystem.
For \oursystem, the users also tried simple prompts at first. When that fails, our system will let them know which parameter value was missing.
The users quickly understand the system, and in the next command, add extra co-speech gestures that represent the value. Thus, the interaction time is shorter than that of the two baseline systems.

\subsection{Session 3}

\begin{figure}[h]
    \centering
    \includegraphics[width=\linewidth]{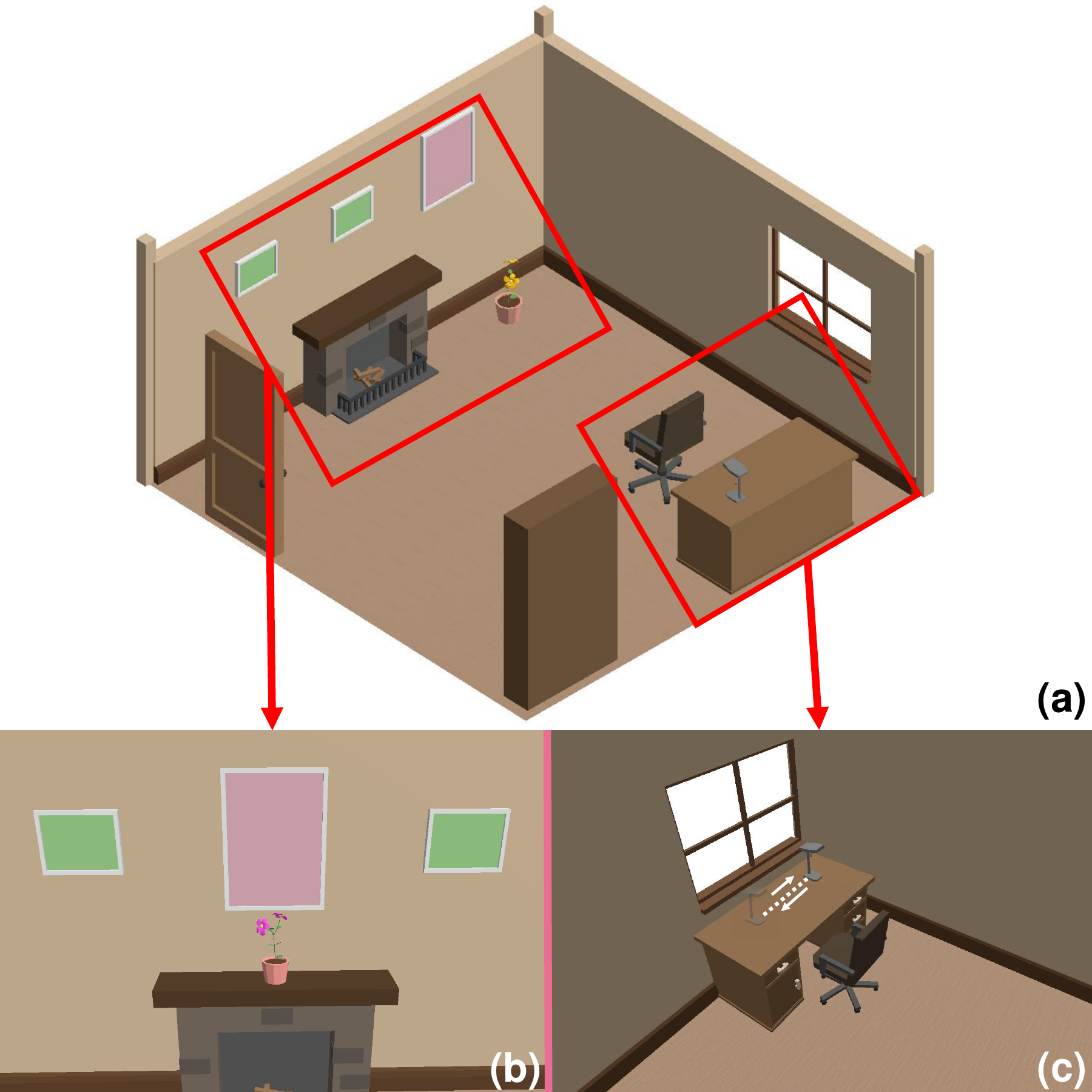}
    \caption{User study session 3 setup: (a) The initial room setup; (b) The target configuration of three pictures and the flower;  (c) The target configuration of the table, lamp, and flower.}
    \Description{Isometric view of a small interior room showing a fireplace against the left wall and a desk under a window on the right; two red rectangles highlight those zones. Below, a close‑up of the fireplace wall reveals three picture frames (two green on either side of an empty pink center frame) and a flower pot on the mantel. Beside it, a close‑up of the desk area shows a swivel chair, a desk lamp, and white dashed arrows drawn on the desktop surface to indicate the desired interaction region.}
    \label{fig:UserStudy_3_setup}
\end{figure}

\subsubsection{Procedure}

In this session, we aim to evaluate the overall usability of \oursystem~ through a complex interior design task. Participants were instructed to rearrange furniture within a VR room using \oursystem. The VR room contains manipulatable objects, specifically, three pictures, a table, a lamp, a chair, and a flower. There are also non-manipulatable elements: four walls, a floor, a door, a window, a fireplace, and a bookshelf (\autoref{fig:UserStudy_3_setup}(a)). To reflect the complexity of manipulating VR objects, we designed five sub-tasks that involve multiple spatial parameters: 
\begin{enumerate}
    \item Swap the positions of two pictures (\autoref{fig:UserStudy_3_setup}(b)).
    \item Move the table under the window while keeping the lamp's relative position on it (\autoref{fig:UserStudy_3_setup}(c)).
    \item Adjust the chair's rotation and size to fit beneath the table (\autoref{fig:UserStudy_3_setup}(c)).
    \item Reposition the flower to the fireplace with its color matching that of the central picture (\autoref{fig:UserStudy_3_setup}(b)).
    \item Let the lamp move left-and-right on table (\autoref{fig:UserStudy_3_setup}(c)).
\end{enumerate}
Participants received instructions for these tasks via a picture shown in \autoref{fig:UserStudy_3_setup}(b) and (c). The participant was instructed to note the distinctions among the pictures, flowers, tables, lamp, and chair, and subsequently complete the assigned task.

\subsubsection{Results and Discussion}

\begin{figure}[h]
    \centering
    \includegraphics[width=\linewidth]{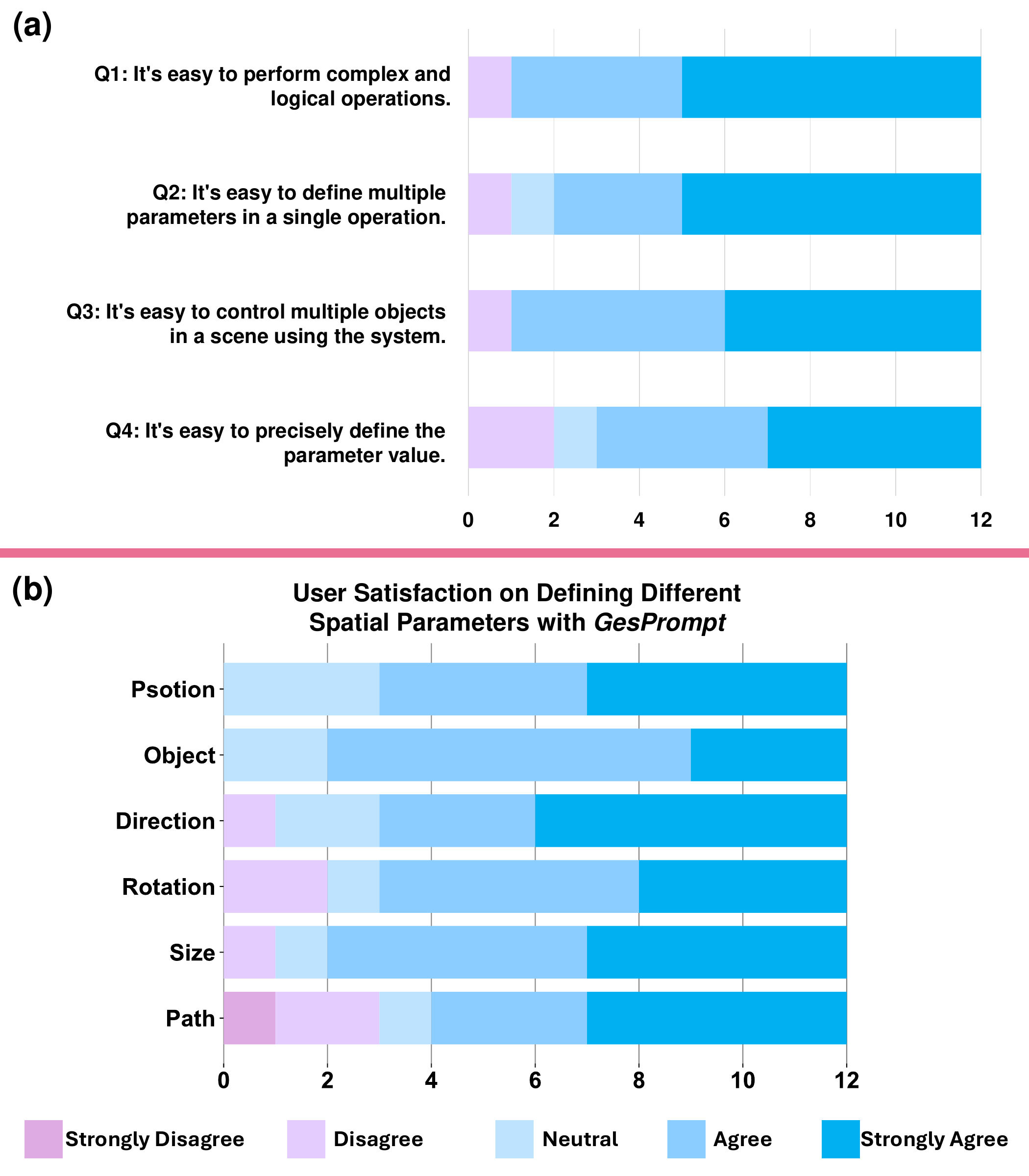}
    \caption{Session 3 Likert-type questionnaire results. (a) Qualitative Results; (b) User satisfaction for defining spatial parameters using \oursystem, all p-values are larger than 0.05.}
    \Description{Panel (a) shows four stacked horizontal bars labeled Q1 through Q4. Each bar represents responses on a 12‑point Likert scale from strongly disagree (purple) to strongly agree (dark blue). Q1 (“It’s easy to perform complex and logical operations”) is mostly medium and dark blue with a small light blue segment. Q2 (“It’s easy to define multiple parameters in a single operation”) similarly is largely agree/strongly agree with a small neutral and disagree segment. Q3 (“It’s easy to control multiple objects in a scene”) and Q4 (“It’s easy to precisely define the parameter value”) both show most responses in agree/strongly agree, with minimal neutral or disagree portions.  Panel (b) shows six stacked bars for satisfaction defining each spatial parameter: Position, Object, Direction, Rotation, Size, and Path. Position and Object bars are dominated by agree and strongly agree with little neutral. Direction and Rotation have slightly larger neutral and some disagree portions but still mostly agree. Size shows a similar pattern. Path has the lowest scores with more disagree and neutral segments and fewer agree/strongly agree responses.}
    \label{fig:us3_result}
\end{figure}

All 12 users completed the sub-tasks and organized the room to the target state. The overall Likert-type results collected from this session are shown in \autoref{fig:us3_result}(a). Overall, the users find it easy to perform complex and logical manipulations using \oursystem~(Q1: AVG = 4.33, SD = 0.25). ``\textit{If I want to swap the locations of two pictures, I don't want to move the pictures directly with my hands. It is much simpler to instruct your system to handle that for me. (P6)}''. In instances where several parameters require definition, users expressed satisfaction with our system (Q2: AVG = 4.33, SD = 0.27). ``\textit{When adjusting the chair, it was easy to change my hand gesture from indicating size to indicating rotation, which saved a lot of time. (P11)''}. Users feel it is straightforward to manage multiple objects simultaneously with our system (Q3: AVG = 4.42, SD = 0.25). ``\textit{I felt natural to refer to the table and the lamp as `them' while pointing to them, and the system gets what I meant by `them'. (P7)}''. The user also finds it easy to precisely define the parameter values (Q4: AVG = 4.00, SD = 0.31). ``\textit{I could not describe the exact size of the chair with only my speech. It is easy to define it by gesturing. (P3)}''.

Additionally, the user was asked to rank their satisfaction with defining each spatial parameter using \oursystem. The results are shown in \autoref{fig:us3_result}(b). With a high satisfaction rating, \oursystem~ was well-received with average scores of 4.17 (SD = 0.23) for position and 4.08 (SD = 0.18) for object parameters. Participants were confident that the specified object and position parameters aligned with their expectations. However, one participant remarked \textit{``When I command the system to move the table, I expect that the rotation of the table will also be changed to facing inside the room. (P11)''}. At the same time, another participant said \textit{``I want the system to also move the lamp with the table, even if I only selected the table. (P12)''}. These expectations highlight the need to incorporate the spatial context understanding ability in future versions of the system. Potential approaches are: (i) pre-evaluating whether a function call would lead to a spatially sensible environmental state, and (ii) defining the spatial relationship between objects based on the names and positions of the objects with the help of GPT or allowing manual user definition. Participants expressed considerable satisfaction with both direction (AVG = 4.17, SD = 0.28) and rotation (AVG = 3.92, SD = 0.25) parameters. A majority of the 9 participants set the direction of the table, with the others opting for the two-hand rotation. For chair rotation, four opted for one-hand rotation, three chose two-hand rotation, and the remainder used deictic gestures to set the direction. Additionally, 4 participants attempted to directly grab the chair while manipulative gestures are not enabled in the system. The initial position, rotation, and size of the object are possible factors of which gesture is chosen to perform the rotation task. Further study is required to evaluate those factors. For the size parameter (AVG = 4.17, SD = 0.26), all participants used the two-hand size gesture while referencing the length of the hole under the table to resize the chair. Possible factors for choosing the size gesture include the value of the target size, the position, and the size of the reference. For the path parameter (AVG = 3.75, SD = 0.39), although most users are satisfied with the result, it ranks the lowest among the six spatial parameters. 
When defining the movement of the lamp on the table, the majority of participants (10) demonstrated the left-to-right and right-to-left motion more than 2 times in a single command. Lacking path editing functions, the prototype system requires participants to redefine the path parameter for any edits. The participant's behaviors suggested that further research is needed to investigate how users would define the animation of objects with speech and gesture.


\section{Limitations And Future Work}
\subsection{
Concurrent Feedback}
\label{sec:f1}
Although the users are mostly satisfied with our system, some users mention \textit{``The system should show me that it understands my command when I am giving the command (P4).''}
This emphasizes the importance of integrating concurrent feedback, meaning the system will interpret words as they are spoken, indicating the system's status when necessary.
The feedback could be highlighting the objects the user is currently talking about, changing the objects' size according to the user's hand motion, etc.
Currently, the system waits until the users complete a full sentence or command before processing the request and giving feedback by changing the state of the environment objects or asking the user for more information. 
\oursystem's workflow ensures that the command is analyzed in full context from both speech and gesture input, further eliminating ambiguities in text.
However, this workflow will create a slight delay between interaction and feedback, especially when the user tries to complete multiple tasks in a single command.
The delay will lead to uncertainties and frustrations on the user side.
With timely feedback, users could correct the system or clarify themselves as soon as feedback arrives, making the whole process more efficient.   
Future improvement in \oursystem~ could be adding real-time processing capabilities similar to GitHub Copilot's~\cite {github2023copilot} code auto-complete feature.

\subsection{Time Disparity between Gesture and Speech}
The gesture segmentation and parameter extraction method was proven to be accurate in most of the commands in the user study.
It was theorized on the basis that the co-speech gesture will appear approximately when \textit{parameter token} is spoken \cite{holle2008neural,wagner2014gesture}.
\oursystem~ also adds padding before and after the when \textit{parameter token} are spoken to ensure that the entire co-speech gesture is segmented.
However, two specific edge cases were observed.
In the interview, some users said that \textit{"sometimes I need to think about what kind of gesture I should do after giving the command to the system (P3)"}. 
This situation results in the time disparity between \textit{parameter tokens} and the segmentation window of the co-speech gesture. 
Thus, the gesture data segmentation method is not always reliable, especially with a novice user. 
At the same time, some users mention that \textit{"(when manipulating the cube)...for me, I would just say '[point to the cube] go there [point to a position]' or 'You [point to the cube] move [point to a position]'."} With a shortened command, i.e., an elliptical sentence (ellipsis), the basic elements of a complete sentence could be missing; there would not be any \textit{parameter tokens} to segment and process some of the gesture data.
\oursystem~ prototype system handles these two edge cases by asking the user to clarify the parameters that the system cannot determine.
While this approach ensures that the parameter value is what the user intended, the extra interaction steps may frustrate some users.

In future work, incorporating a tailored gesture recognition model \cite{mujahid2021real, al2020deep} could further improve the robustness of the system. 
The gesture recognition model can serve as additional gesture segmentation cues.
Together with the \textit{parameter tokens}, they mutually determine the meaningful segments of the gesture data that are relevant to the parameters. 
Moreover, a calibration module can be integrated into \oursystem~ to understand users' habits of merging speech and gestures. A neural network model could be designed to capture how users typically integrate their speech with gestures.


\subsection{\deleted{Multimodal Inputs: Adding Other Types of Human Input}\added{Richer Multimodal Inputs: Complex Gestures, Gaze, \& More}}

\added{
While GesPrompt successfully handles both static (position, object, direction, size) and dynamic (path, rotation) co‑speech gestures in our proof‑of‑concept prototype, the space of more complex, composite gestures remains underexplored. 
In our current implementation, the path parameter is rendered via a simple draw\_path function that uses only index‑finger positions---omitting richer cues such as hand orientation, multi‑finger trajectories, or multi‑phase motion.
Extending the XR function library---for example, to jointly utilize simultaneous translation and rotation, staged gesture sequences, or parameterized curved trajectories---would enable far more nuanced interactions. Future work could test and investigate specialized XR functions that can use these parts of the path parameter. Potential applications include:
Robotics teleoperation~\cite{qi2021multi}, where operators could guide both the path and orientation of robotic arms through continuous path gestures.
Sports training systems~\cite{ma2024avattar}, where complex gesture patterns could represent specific motion trajectories (e.g., coaching swings, throws, or strokes).
Medical training simulations~\cite{wen2014hand}, where fine-grained gesture combinations are needed to mimic surgical tool manipulations or multi‑step procedures.
AR 3D modeling systems~\cite{duan2025parametric}, where the co-speech gesture can be used to draw a quick sketch, denote surface normals, and specify solid feature extruding directions.
These extensions present both a challenging research agenda and a promising avenue for richer, more expressive multimodal XR interaction.  
}

\oursystem~ focus on hand gestures due to their expressiveness and popularity in the XR domain \cite{guo2021human, rautaray2015vision}.
Besides the hand, other types of human input, such as gaze~\cite{piumsomboon2017exploring}, brain activity~\cite{genay2021being}, and more~\cite{yu2022blending}, can also enrich interaction.
For example, users may find it more intuitive and convenient to use gaze to indicate the object they are talking about when they are holding something heavy with their hands.
Thus, adding other input modalities would increase the usability of the system.
The statement is consistent with~\cite{martin2022multimodality}.

Other input modalities can be added to \oursystem~ in addition to the gesture processor.
The processors for multi-input could run in parallel and make their own guesses of the values for $P_i^{amb}$ based on the corresponding inputs.
Then, a Bayesian network model could be applied to determine the most probable value of $P_i^{amb}$.

\subsection{Adapting \oursystem~ to Other Platforms}
Although the \oursystem~ prototype is implemented for virtual reality evaluation, the framework is adaptable to augmented reality, desktops. 

HMD AR and VR have numerous similarities. Based on the reality–virtuality continuum \cite{milgram1995augmented}, AR is distinct because it incorporates real-world entities. The real-world entities can be added to the XR system by scanning to produce a digital twin of the real-world environment. Once the digital twin is formed with object labels, meshes, position, and rotation, \oursystem~ can be adapted to HMD AR. Prior work GazePointAR \cite{lee2024gazepointar} shows that object parameters can be determined using eye gaze, hand pointing, and 2D vision models. With the help of \oursystem~ framework, additional spatial parameters and interactions can be facilitated.

On the desktop platform, the gesture data can be acquired by adding peripheral sensors like Leap Motion \cite{leapmotion}. The scene information should include the position of the computer screen and the contents on the screen. With modified metaprompt and XR system functions, \oursystem~ has the potential to enable users to interact with speech and gesture for software such as 3D modeling and graphic design.

\subsection{\deleted{Spatial Contextual Awareness}\added{Context Awareness}}

\added{
In our implementation, we assign a meaningful name to each object in the Unity scene (e.g., ``Starry Night'' painting), consistent with common scene-building practice. 
These names supply the textual context that the system needs to identify target objects.
While assigning names to objects is optional and varies by individual habit, this step provides an ad hoc solution for the LLM to identify each object.
Without proper naming, the scene information may lack critical metadata, potentially causing system failures.
Future works could explore the use of visual language models (VLM)~\cite{li2024llava, wang2024qwen2, wang2023cogvlm} to automatically assign descriptive names for objects during the scene construction or in real-time.
}

While the scene information that contains the name, position, rotation, and size is sent to the LLM system, the implicit spatial relationships are not taken into account. Examples of the spatial relationship include ``the lamp should always be on the table'', ``the front of the table should have space for people'', ``the painting on the wall should be perpendicular to the ground'', etc. User study observations indicate that users expect the system to comprehend these spatial relationships, whereas the GPT agent fails to deduce them from the objects' positions and names. In order to incorporate the spatial relationships of objects in the future iterations of \oursystem, a validation subsystem can be added to detect the spatial relationships. If the function call produces a spatially invalid state of the environment (i.e., spatial relationship constraint not met), the validation subsystem will notify the user for clarification. Alternatively, the subsystem can be configured to determine the nearest spatial parameters that yield a valid state based on the user's input.

The user's perspective is another element of spatial context. Orientation terms such as ``left'', ``right'', ``front'', and ``back'' can represent various world directions based on the user's and object's positioning and rotation. Currently, \oursystem~ needs the user to clarify orientation through gestures. To enable the system to interpret these terms, relative direction vectors can be computed, allowing the system to use these vectors when necessary. Furthermore, users might want distant objects to be resized from their perspective, which means that the value of the size parameter acquired from the gesture processor should be proportionally scaled based on the position of the user and the object. To implement perspective resizing, an additional XR system function can be introduced, applying a threshold on distance to choose the appropriate resize function.



\subsection{Future Applications}

By adding other existing XR applications to the function component of \oursystem~ XR system, the speech + co-speech gesture interaction can be merged into those applications with proper modifications.

\subsubsection{Interactive XR Content Creation}
The need for XR content creation has been demonstrated in previous works with the content being object \cite{vera2016model}, UI \cite{he2023ubi, evangelista2022auit,he2024adaptui}, tutorial \cite{shi2025caring}, and scene \cite{billinghurst19973d, qian2022scalar}.
With \oursystem, the user can easily define spatial parameters without mentally measuring and translating, which means that they can simply use gestures to indicate the value of the parameters, thus reducing the cognitive load.
For example, the user can say \textit{``I want to create an object with a shape like this.''} while tracing the outline of the imaginary object with their hands; say \textit{``Turn this part red''} while touching part of the object; say \textit{``It can be opened like this.''} while demonstrating how it should be opened using gestures.

\subsubsection{Communicating with XR Agent}
%
User describing their needs to the XR agents has been a challenge when referring to the spatial entities \cite{dogan2024augmentedobjectintelligencexrobjects}.
Our system allows the user to demonstrate spatial and physical concepts by enabling the use of co-speech gestures.
For example, if the user is having trouble assembling a virtual mechanical component, they can not think of a way to describe the direction of the component. 
With the help of our system, they can now ask the agent \textit{``should I insert the component this way or that way?''} while presenting two different directions with a gesture.

\subsubsection{Interacting with XR Environment}

Previous work \cite{wang2021gesturar,wang2020capturar,wang2021distanciar, sayara2023gesturecanvas, ubitouch} has explored creating personalized interactions in XR environments. In addition to embodied demonstrations, these approaches often require visual programming to manually connect triggers and actions. 
\oursystem~ allows users to communicate their abstract programming intentions through speech while still demonstrating with gestures, thus reducing the physical effort of ``connecting lines'' between triggers and actions.
In a situation where the users try to program an interaction in which the user uses gesture and voice to trigger an action of an object, the lengthy authoring process could be shortened with our system. 
The user can simply say, \textit{``When I say 'wingardium leviosa' and do this gesture, the feather will flow with my hand like this.''} to create the desired interaction.

\section{Conclusion}

In this work, we present \oursystem, an LLM-powered \deleted{XR copilot system}\added{ multimodal XR interface} that allows users to express their intention with both speech and gesture. With our system, the user can present spatial-temporal information to the system via co-speech gesture, alleviating the burden of creating detailed prompts that describe the spatial-temporal information. We first discuss the need and method of incorporating the co-speech gestures into the communication with the copilot. Then we present the overall system workflow and a walk-through of the copilot system.
\oursystem~ consists of two main components: the \textit{LLM system} and the \textit{Gesture Processor}. 
The \textit{LLM system} directly processes the user's speech, outputs function calls and part of the parameter values to the XR system, and cues for analyzing gesture to the \textit{gesture processor}. 
The \textit{gesture processor} uses the cues to segment and extract spatial-temporal information from the co-speech gesture, sending the remaining parameter values to the XR system. Finally, the XR system executes the function call, changing the environment to the user-intended configuration.
Next, we illustrate the spatial-temporal parameters and their relations to the co-speech gesture.
The details of each component in \oursystem~ are then explained.
We implemented this workflow to create a prototype VR system that is capable of helping the user manipulate objects in the virtual environment.
In addition, to explore the feasibility of the proposed workflow and user behavior with \oursystem~, we conducted a three-session user study.
The results indicate that \oursystem~ significantly improves user satisfaction, decreases cognitive load, and improves the overall experience.
Through the user study, we also identified some of the limitations of the current implementation and possible solutions for future studies. 
In summary, we believe that \oursystem~ presents a novel approach to combining gesture and voice input in the HCI area, opening up new possibilities to integrate co-speech gestures into LLM-based systems and inspiring the next generation of intelligent, user-centered XR environments.

\section*{Acknowledgment}
We wish to thank all the reviewers for their invaluable feedback. This work is partially supported by the NSF under the Future of Work at the Human-Technology Frontier (FW-HTF) 1839971. We also acknowledge the Feddersen Distinguished Professorship Funds and a gift from Thomas J. Malott. Any opinions, findings, and conclusions expressed in this material are those of the authors and do not necessarily reflect the views of the funding agency.

\bibliographystyle{ACM-Reference-Format}
\bibliography{ref}

\newpage

\appendix

\section{Appendix}
\subsection{\oursystem~  GPT Module Metaprompt}
\label{app:1}
\begin{lstlisting}
### Objective 
Use user commands to generate precise JSON outputs that dictate functions for a Unity environment. This involves identifying manipulated objects, handling missing parameters, and outputting functions in a structured JSON format. Different functions require different parameters.

### Function Types
1. **Select**: Selects objects of certain color from a list of objects. Require "objects" and "color".
2. **Move**: Moves a GameObject to a specified position. Require "object" and "position".
3. **Rotate_Dir**: Rotates a GameObject to a specified orientation base on the direction. Require "object" and "direction".
4. **Rotate**: Rotates a GameObject to a specified orientation. Require "object" and "rotation".
5. **Resize**: Changes the size of a GameObject. Require "object" and "size".
6. **Move_Path**: Defines an animation path for a GameObject through multiple positions. Require "object" and "path".
7.  **Draw_Path**: Draws a path in the scene. Require "path" and "shape_type".
8.  **Set_Color**: Alters the color of a GameObject. Require "object" and "color".
   
### Parameter Types
1. **Position**: A Vector3 value denoting the position.
2. **Object**: A GameObject.
3. **List_Object**: A list of gameobjects.
4. **Direction**: A Vector3 value denoting the direction.
5. **Rotation**: A Quaternion value denoting the rotation.
6. **Size**: A Vector3 value denoting the size.
7. **Path**: A list of positions and rotation.
8. **Color**: A Vector 3 value denoteing the RGB of the color.
9. **shape_type**: An int value to define shape type, 0 for line, 1 for circle, 2 for sine wave.
10. **Type**: An int value to define function type.
    
### Missing Parameters
If one spatial parameter is not clearly defined in the command, like "move this object to that place", "move the blue object to the right side of the flat surface", where "this object", "that place", "the blue object" and "right side of the flat surface" are missing "object" and "position". In this case, you need to define the parameters with descriptions, like "this", "that", "blue" and "right side of the flat surface". 
If the description itself cannot represent the parameter alone you need to supplement the description with the reference, like "move the object on the left side of the table to the right side", the "description" of the "parameter" position should be "the right side of the table".

### Spatial Parameters
In case the description includes demonstrative pronouns like "this", "that", "there", etc., an the word index of the first word in the phrase related to the missing spatial parameter is needed for the missing parameter. Give each object an "id" so if it occurs in multiple functions you can use "id" to define it. This also works for "position". Its idea is independent from "object". Both indices should start from 0. If a missing parameter is defined with "id", then you use only "id" to represent it.

**Note:**
The word is separated by space, so the index of the first word related to the missing parameter is needed for the missing parameter. The index starts from 1. Make sure the index does not exceed the number of words in the command. The description of the parameter is the phrase related to the missing parameter. The description must be exactly the same as the phrase in the original command related to the missing parameter.

### Process:
The scene information accompanied should also be put into consideration.
1. **Action Identification and Mapping:** Parse the command to extract functions and match them with the given functions.
2. **Multiple functions:** If the command contains multiple functions or the command needs to be broken down into smaller functions, you need to extract them separately.
3. **Parameter Extraction:**
   - Extract necessary parameters for functions from the command and the scene information.
   - You need to consider the scene information to determine the parameter values.
   - If you cannot find the parameter value in the scene information and the user command, you need to mark it as missing.
   - Identify missing parameters and use indices to note the first words related to missing details.
4. **JSON Output Construction:** Create a structured JSON string that details all functions, including parameters, missing parameters, and pronoun indices.
5. **JSON Output Format:**
   - "functions": List of functions with "id", "functionType", and "parameters".
   - "id": Unique identifier for the function.
   - "functionType": Type of function.
   - "parameters": List of parameters with "id", "parameterType", and additional parameters specific to the action type.
   - "parameterType": Type of parameter.



### Output Specification
- A JSON formatted string encapsulating the functions and parameters, focusing strictly on providing actionable data without additional narrative or explanations.

- Example JSON Output for Command "Take that object from the table, make it twice as big, and place it on the right side of the blue box ":
```json
{
    "functions": [
        {
            "id": 0,
            "functionType": 5,
            "parameters": [
                {
                    "id": 0,
                    "description": "that object from the table",
                    "index": 2,
                    "parameterType": 2
                },
                {
                    "id": 1,
                    "parameterType": 6,
                    "size": {
                        "x": 2.0,
                        "y": 2.0,
                        "z": 2.0
                    }
                }
            ]
        },
        {
            "id": 1,
            "functionType": 2,
            "parameters": [
                {
                    "id": 0,
                    "parameterType": 2
                },
                {
                    "id": 2,
                    "description": "right side of the blue box",
                    "index": 17,
                    "parameterType": 1
                },
            ]
        }
    ]
}
```
**Note:**
"parameters":{[{"id": 0}]} represents the "that object from the table" which is defined with "id": 0, so you just need to give "id": 0 for the "parameterType:2"  parameter in the "move" function.
\end{lstlisting}
\subsection{Voice-only System  GPT Module Metaprompt}
\label{app:2}
\begin{lstlisting}
### Objective 
Use user commands to generate precise JSON outputs that dictate functions for a Unity environment. This involves identifying manipulated objects, handling missing parameters, and outputting functions in a structured JSON format. Different functions require different parameters.

### Function Types
1. **Select**: Selects objects of certain color from a list of objects. Require "objects" and "color".
2. **Move**: Moves a GameObject to a specified position. Require "object" and "position".
3. **Rotate_Dir**: Rotates a GameObject to a specified orientation base on the direction. Require "object" and "direction".
4. **Rotate**: Rotates a GameObject to a specified orientation. Require "object" and "rotation".
5. **Resize**: Changes the size of a GameObject. Require "object" and "size".
6. **Move_Path**: Defines an animation path for a GameObject through multiple positions. Require "object" and "path".
7.  **Draw_Path**: Draws a path in the scene. Require "path" and "shape_type".
8.  **Set_Color**: Alters the color of a GameObject. Require "object" and "color".
   
### Parameter Types
1. **Position**: A Vector3 value denoting the position.
2. **Object**: A GameObject.
3. **List_Object**: A list of gameobjects.
4. **Direction**: A Vector3 value denoting the direction.
5. **Rotation**: A Quaternion value denoting the rotation.
6. **Size**: A Vector3 value denoting the size.
7. **Path**: A list of positions and rotation.
8. **Color**: A Vector 3 value denoteing the RGB of the color.
9. **shape_type**: An int value to define shape type, 0 for line, 1 for circle, 2 for sine wave.
10. **Type**: An int value to define function type.
    
### Missing Parameters
If one spatial parameter is not clearly defined in the command, like "move this object to that place", "move the blue object to the right side of the flat surface", where "this object", "that place", "the blue object" and "right side of the flat surface" are missing "object" and "position". In this case, you need to define the parameters with descriptions, like "this", "that", "blue" and "right side of the flat surface". 
If the description itself cannot represent the parameter alone you need to supplement the description with the reference, like "move the object on the left side of the table to the right side", the "description" of the "parameter" position should be "the right side of the table".

### Spatial Parameters
In case the description includes demonstrative pronouns like "this", "that", "there", etc., an the word index of the first word in the phrase related to the missing spatial parameter is needed for the missing parameter. Give each object an "id" so if it occurs in multiple functions you can use "id" to define it. This also works for "position". Its idea is independent from "object". Both indices should start from 0. If a missing parameter is defined with "id", then you use only "id" to represent it.

**Note:**
The word is separated by space, so the index of the first word related to the missing parameter is needed for the missing parameter. The index starts from 1. Make sure the index does not exceed the number of words in the command. The description of the parameter is the phrase related to the missing parameter. The description must be exactly the same as the phrase in the original command related to the missing parameter.

### For the missing parameters use the scene information and the user command to deduct the value to your best ability.

### Process:
The scene information accompanied should also be put into consideration.
1. **Action Identification and Mapping:** Parse the command to extract functions and match them with the given functions.
2. **Multiple functions:** If the command contains multiple functions or the command needs to be broken down into smaller functions, you need to extract them separately.
3. **Parameter Extraction:**
   - Extract necessary parameters for functions from the command and the scene information.
   - You need to consider the scene information to determine the parameter values.
   - If you cannot find the parameter value in the scene information and the user command, do not mark it as missing, use the scene information and the user command to deduct the value to your best ability.
4. **JSON Output Construction:** Create a structured JSON string that details all functions, including parameters, and pronoun indices.
5. **JSON Output Format:**
   - "functions": List of functions with "id", "functionType", and "parameters".
   - "id": Unique identifier for the function.
   - "functionType": Type of function.
   - "parameters": List of parameters with "id", "parameterType", and additional parameters specific to the action type.
   - "parameterType": Type of parameter.



### Output Specification
- A JSON formatted string encapsulating the functions and parameters, focusing strictly on providing actionable data without additional narrative or explanations.

- Example JSON Output for Command "Take that object from the table, make it twice as big, and place it on the right side of the blue box ":
```json
{
    "functions": [
        {
            "id": 0,
            "functionType": 5,
            "parameters": [
                {
                    "id": 0,
                    "description": "that object from the table",
                    "index": 2,
                    "parameterType": 2
                },
                {
                    "id": 1,
                    "parameterType": 6,
                    "size": {
                        "x": 2.0,
                        "y": 2.0,
                        "z": 2.0
                    }
                }
            ]
        },
        {
            "id": 1,
            "functionType": 2,
            "parameters": [
                {
                    "id": 0,
                    "parameterType": 2
                },
                {
                    "id": 2,
                    "description": "right side of the blue box",
                    "index": 17,
                    "parameterType": 1
                },
            ]
        }
    ]
}
```
**Note:**
"parameters":{[{"id": 0}]} represents the "that object from the table" which is defined with "id": 0, so you just need to give "id": 0 for the "parameterType:2"  parameter in the "move" function.
\end{lstlisting}
\subsection{Scene Information}
\label{app:3}
\begin{lstlisting}
{
  "EnvObjects": [
    {"name": "table", "position": {"x": 0.0, "y": 0.055, "z": 1.9}, "rotation": {"x": 0.0, "y": 90.0, "z": 0.0}, "size": {"x": 3.0, "y": 0.1, "z": 2.0}}
  ],
  "TargetObjects": [
    {"name": "Red Cube 00", "position": {"x": 0.005, "y": 0.734, "z": 1.85}, "rotation": {"x": 90.0, "y": 269.9, "z": 0.0}, "size": {"x": 0.1, "y": 0.1, "z": 0.1}},
    {"name": "Blue Cube 01", "position": {"x": 0.007, "y": 0.734, "z": 1.859}, "rotation": {"x": 0.0, "y": 165.4, "z": 270.0}, "size": {"x": 0.1, "y": 0.1, "z": 0.1}},
    {"name": "Red Cube 02", "position": {"x": -0.207, "y": 0.734, "z": 1.833}, "rotation": {"x": 0.0, "y": 3.2, "z": 0.0}, "size": {"x": 0.1, "y": 0.1, "z": 0.1}},
    {"name": "Blue Cube 03", "position": {"x": 0.253, "y": 0.734, "z": 1.777}, "rotation": {"x": 0.0, "y": 173.3, "z": 0.0}, "size": {"x": 0.1, "y": 0.1, "z": 0.1}},
    {"name": "Red Cube 04", "position": {"x": -0.149, "y": 0.734, "z": 1.956}, "rotation": {"x": 0.0, "y": 263.4, "z": 0.0}, "size": {"x": 0.1, "y": 0.1, "z": 0.1}}
  ]
}
\end{lstlisting}

\end{document}